\documentclass[journal]{IEEEtran}
\usepackage{cite}
\usepackage[table]{xcolor}
\usepackage{amsmath,amssymb,amsfonts}
\usepackage{graphicx,color}
\usepackage{textcomp}
\usepackage{caption}
\usepackage{wrapfig}
\usepackage{lipsum}
\usepackage[hidelinks]{hyperref}
\usepackage{flushend}
\usepackage{setspace}
\usepackage{booktabs}
\usepackage{makecell}
\usepackage{comment}
\usepackage{subcaption}
\usepackage{mathtools}
\usepackage{amsthm}
\usepackage{bbold}
\usepackage[nopostdot,acronym,shortcuts,nonumberlist]{glossaries}

\usepackage{tikz}
\usetikzlibrary{decorations.pathreplacing,positioning, arrows.meta}
\usepackage[linesnumbered, ruled,vlined]{algorithm2e}
\newenvironment{fivept}
  {\begingroup\fontsize{8pt}{10pt}\selectfont}
  {\endgroup}

\newtheorem{definition}{Definition}[]

\newtheorem{example}[definition]{Example}

\definecolor{tudablue}{HTML}{000000}
\newcommand{\tuda}[1]{\textcolor{black}{#1}}
\newcommand{\rc}[1]{\textcolor{black}{#1}}

\newcommand{\rev}[1]{\textcolor{tudablue}{#1}}

\def\sysname{\textsc{PSBO}\xspace}
\def\valname{data quality\xspace}
\def\valindex{qual}

\begin{document}

\pagestyle{empty}

\vspace*{3.5cm}
\hspace{0.2cm}
\fboxrule1.5pt
\framebox[17cm]{
\parbox{16cm}{
\vspace{0.3cm}

\copyright 2026 IEEE.  Personal use of this material is permitted.  Permission from IEEE must be obtained for all other uses, in any current or future media, including reprinting/republishing this material for advertising or promotional purposes, creating new collective works, for resale or redistribution to servers or lists, or reuse of any copyrighted component of this work in other works.

\vspace*{0.3cm}
}}

\title{\fontsize{28pt}{32pt}\selectfont 
Deep Sleep Scheduling for Satellite IoT via Simulation-Based Optimization
}

\author{Wanja de Sombre, Monika Tomová, Marek Galinski, Anja Klein, Andrea Ortiz
\thanks{Wanja de Sombre (w.sombre@nt.tu-darmstadt.de) and Anja Klein (a.klein@nt.tu-darmstadt.de) are with the Communications Engineering Lab, Technische Universität Darmstadt, Germany. Monika Tomová (m.tomova@protonmail.com) and Marek Galinski (marek.galinski@stuba.sk) are affiliated with the Automotive Innovation Lab, Slovak University of Technology in Bratislava, Slovakia. Andrea Ortiz (andrea.ortiz@tuwien.ac.at) is with the Institute of Telecommunications, Vienna University of Technology, Austria.}
\thanks{This work was funded by the BMFTR project Open6GHub+ under grant 16KIS2407, by DAAD with funds from the German Federal Ministry of Research, Technology and Space BMFTR under grant 57817830, and by the LOEWE Center emergenCITY under grant LOEWE/1/12/519/03/05.001(0016)/72. The work of Andrea Ortiz was funded by the Vienna Science and Technology Fund WWTF Grant ID 10.47379/VRG23002.}}

\maketitle

\newacronym{aoi}{AoI}{Age of Information}
\newacronym{mse}{MSE}{Mean Squared Error}
\newacronym{aoii}{AoII}{Age of Incorrect Information}
\newacronym{aos}{AoS}{Age of Synchronisation}
\newacronym{qaoi}{QAoI}{Query Age of Information}
\newacronym{paoi}{PAoI}{Peak Age of Information}
\newacronym{iot}{IoT}{Internet of Things}
\newacronym{siot}{S-IoT}{Satellite Internet of Things}
\newacronym{iiot}{IIoT}{Industrial Internet of Things}
\newacronym{mdp}{MDP}{Markov Decision Process}
\newacronym{sus}{SUS}{Status Update System}
\newacronym{ack}{ACK}{acknowledge}
\newacronym{nack}{NACK}{negative acknowledge}
\newacronym{got}{GoT}{Goal-oriented Tensor}
\newacronym{ble}{BLE}{Bluetooth Low Energy}
\newacronym{leo}{LEO}{Low Earth Orbit}
\newacronym{harq}{HARQ}{hybrid automatic repeat request}
\newacronym{fdpg}{FDPG}{Finite Difference Policy Gradient}

\nocite{
Wang2024_2, 
Cao2023, 
Gong2022, 
Badia2025, 
Dongare2024, 
Chen2021, 
Chen2024, 
Kam2020, 
deSombre2025b, 
Kriouile2022, 
Maatouk2022, 
deSombre2023b, 
Hong2024, 
Yan2024, 
Gursoy2024}

\begin{abstract}
The Satellite Internet of Things (S-IoT) enables global connectivity for remote sensing devices that must operate energy-efficiently over long time spans.
We consider an S-IoT system consisting of a sender-receiver pair connected by a data channel and a feedback channel and capture its dynamics using a Markov Decision Process (MDP). 
To extend battery life, the sender has to decide on deep-sleep durations. 
Deep-sleep scheduling is the primary lever to reduce energy consumption, since sleeping devices consume only a fraction of their idle power.
By choosing its deep-sleep duration online, the sender has to find a trade-off between energy consumption and \valname degradation at the receiver, captured by a weighted sum of costs.
We quantify \valname degradation via the recently introduced Goal-Oriented Tensor (GoT) metric, which can take both age and content of delivered data into account.
We assume a Markovian observed process and Markov channels with time-varying delay and erasure rates. 
The challenge is that content awareness of the GoT metric makes periodic transmissions inherently inefficient. 
Additionally, optimal sleep durations depends on the (unknown) future states of the observed process and the channels, both of which must be inferred online. 
We propose a novel algorithm using probabilistic simulation-based optimization (PSBO). 
With PSBO, the sensor forecasts future states based on estimated transition probabilities, and uses these forecasts to select the optimal deep-sleep duration. 
\rev{Extensive simulations demonstrate the strong performance of \sysname across diverse conditions.
In S-IoT hardware experiments, \sysname reduces costs by $59\%$ versus a threshold-based solution and by $89\%$ versus Q-learning.}
\end{abstract}

\begin{IEEEkeywords}
Age of Incorrect Information, Satellite Internet of Things, Status Update Systems, Goal Oriented Tensor, Deep Sleep Scheduling
\end{IEEEkeywords}

\section{INTRODUCTION}
\label{sec:intro}
\begin{table*}[t]
    \centering
    \renewcommand{\arraystretch}{1.3}
    \rc{\begin{tabular}{|c|c|c|c|c|c|c|}
        \hline
        \rule{0pt}{0.65cm}
         Reference & \makecell{Age-Based Metric} &  \makecell{Deep-Sleep\\ Scheduling} & \makecell{Content-Aware\\ Metric} &\makecell{Minimization of\\Energy Consumption} & \makecell{Satellite IoT} &  \rc{Years} \\
        \hline
        \hline
        \cite{Wang2024_2, Cao2023, Gong2022} & AoI & \checkmark & & \checkmark & & \rc{2022 - 2024} \\
        \hline
        \cite{Badia2025} & AoI & \checkmark & & \checkmark & \checkmark & \rc{2025} \\
        \hline
        \cite{Dongare2024, Chen2021, Chen2024, Kam2020, deSombre2025b} & AoII & & \checkmark & & & \rc{2020 - 2025} \\
        \hline
        \cite{Kriouile2022, Maatouk2022, deSombre2023b} & AoII & & \checkmark & \checkmark & & \rc{2022 - 2023} \\
        \hline
        \cite{Hong2024} & AoII & & \checkmark & & \checkmark & \rc{2024} \\
        \hline
        \cite{Yan2024} & AoII & & \checkmark & \checkmark & \checkmark & \rc{2024} \\
        \hline
        \cite{Gursoy2024} & AoII & \checkmark & \checkmark & \checkmark & & \rc{2024} \\
        \hline
        \rowcolor[gray]{0.9}
        This work & \cellcolor{red!10}GoT metric & \checkmark & \checkmark & \checkmark & \checkmark & \rc{2026}  \\
        \hline
    \end{tabular}}
    \vspace{0.2cm}
    \caption{Summary of related work.}
    \label{tab:sota}
\end{table*}

\IEEEPARstart{T}{he} \gls{siot} has emerged as a solution for enabling the connectivity of sensors even in remote and infrastructure-sparse regions, where conventional \gls{iot} solutions are unavailable. 
Its potential extends to a wide range of applications, such as agriculture, environmental monitoring, and the support of smart grids \cite{Li2020, Yan2024, Sohraby2018, Ramachandran2024}.
These domains often span sparsely populated areas with intermittent or nonexistent terrestrial coverage, making \gls{siot} the only practical means of maintaining continuous, wide-area sensor connectivity.


Sensors deployed in such remote locations are typically battery-powered, and it is desirable for them to operate autonomously for as long as possible. 
This creates a fundamental tradeoff between their primary function, namely delivering high quality data, and the goal of conserving energy.

A central strategy for energy-efficient \gls{iot} devices is the optimization of deep-sleep cycles (cf. \cite{Cao2023}).
In deep-sleep mode, sensors use only a fraction of the energy they consume when active or idling.
Therefore, optimizing deep-sleep durations is the primary lever for reducing energy consumption.

At the same time, the device aims to minimize \valname degradation at the receiver. 
With \valname, we mean the application specific usefulness of the data available at the receiver. Historically, \valname degradation has often been quantified using the \gls{aoi}, which was introduced in \cite{Kaul2012} and has since received significant academic attention (cf. \cite{Kadota2018}).

The \gls{aoi} only captures data freshness.
Hence, when the \gls{aoi} is applied to an application whose measured values remain constant for extended periods and change only sporadically in abrupt jumps, the same data may be transmitted repeatedly, incurring avoidable energy consumption.

This is avoided by using the \gls{aoii} \cite{Maatouk2020} as a metric for \valname degradation, which measures the time since the receiver had correct information about the observed process for the last time.

More recently, the \gls{got} metric proposed by Li et al. in \cite{Li2024} offers a further generalization, which we use in this work to measure \valname degradation.
The \gls{got} metric is a function of the value of one basic \valname degradation metric like either the \gls{aoi} or the \gls{aoii}, the current state of the observed process and the information at the receiver.
Each combination of these three input values is assigned an application-specific cost. 

For example, in a safety-critical system, a critical state at the observed process that remains unknown to the receiver for an extended period incurs a substantially higher cost than a less severe inconsistency or the same inconsistency over a shorter duration.

Together, these costs form the \emph{goal-oriented tensor}.
In contrast to the \gls{aoi}, the \gls{aoii} and the \gls{got} metric take the content of transmitted data into account.
Therefore, we call the latter metrics \emph{content-aware}.

In our scenario, the \gls{iot} device has to find a tradeoff between energy consumption and \valname degradation. Energy consumption and \valname degradation are both associated with a cost. The \valname degradation cost is measured via the \gls{got} metric and the energy consumption cost is measured in joules. We then minimize a weighted sum of both costs.
\rev{The respective weights account for the relative importance of energy consumption and data quality degradation at the receiver, thereby converting this tradeoff into a single scalar optimization objective.}

When optimizing the described tradeoff between energy consumption and \valname degradation in \gls{siot}, solutions have to take the device's uncertainty about its environment into account. 
The exact behavior of the observed process as well as channel characteristics are a priori unknown at the device.

Here, channel characteristics differ from those in terrestrial \gls{iot}. 
The end-to-end \gls{siot} path spans multiple time-varying links (sensor to satellite terminal, satellite terminal to satellite, inter-satellite, satellite to ground station, and ground stations to server).
Additionally, \gls{leo} satellite constellations are highly dynamic and continually change link geometry and visibility. 
Therefore, path loss, delay, and outage probability can fluctuate strongly over time \cite{Heine2023, Li2020, Kim2024}.

We consider a battery-powered \gls{iot} device that observes a process such as electric current, temperature, acceleration, humidity or light intensity, and transmits measurements to a remote receiver via a satellite link.

Time is modeled in discrete steps, during which the device can either be awake or in deep-sleep mode. 
When awake, it senses the current environment and transmits the obtained measurements.
New measurements are only transmitted if the transmission improves the \valname at the receiver.

\rev{Since both the data and the feedback channel qualities exhibit temporal dependence due to changing link geometry and visibility in the \gls{leo} constellation, we use a Markov chain as a tractable way to model the channel.
We therefore model the time-varying channels as Markov channels.}
With this model, we assume that the channels are packet erasure channels, where erasure rate and delay change according to a Markov process.

To reduce energy consumption, the device may enter a deep-sleep phase for multiple time steps, during which it is entirely inactive and cannot respond to acknowledgments or changes in the environment.
The central decision of the \gls{iot} device is whether to enter deep sleep and, if so, for how long.
When the expected changes in the measured values are moderate, longer deep-sleep durations are possible.
In contrast, high anticipated fluctuations in measurements require shorter deep-sleep intervals.

We propose the \sysname approach, a Probabilistic Simulation-based Optimization (cf. \cite{Amaran2016}):
For each deep-sleep duration, in ascending order, the \emph{probability} of every possible time-averaged cost is estimated by using a \emph{simulation} of the system during deep-sleep.
The device then uses these estimates to calculate expected time-averaged costs for the deep-sleep durations and to \emph{optimize} the deep-sleep duration by choosing the one with the lowest expected time-averaged cost.

To carry out these simulations, the device is dependent on knowledge about the environment, e.g. about the observed process and the channels.
It obtains this knowledge by observing the observed process and the channels over time and thereby gradually improves the simulations of different deep-sleep durations.

Our approach exploits two problem-specific properties of deep-sleep duration selection:

First, instead of a sample-based approach with many simulations, we use a distribution-based simulation. 
Rather than generating many samples and averaging their outcomes, we propagate belief distributions over the costs and compute expected time-averaged costs directly from these distributions.
This enables a substantially more efficient evaluation of candidate deep-sleep durations.

Secondly, we exploit the fact that the distribution-based simulation for a deep-sleep duration of $n+1$ steps can build directly on the distribution-based simulation for a deep-sleep duration of $n$ steps.
Consequently, evaluating a sequence of candidate deep-sleep durations does not require restarting the simulation from scratch for each duration, but can be carried out via a single progressively extended simulation.

The contributions of this paper are as follows:
\begin{itemize}
    \item \rev{To the best of our knowledge, this work is the first to characterize and find a tradeoff between \valname degradation, quantified by the \gls{got} metric, and energy consumption via deep-sleep scheduling in the \gls{siot}.}
    \item \rev{It is also the first to apply probabilistic simulation-based optimization in this context by exploiting the problem-specific structure of the deep-sleep scheduling task to enable an efficient implementation.}
    \item We formulate the problem of energy-efficient deep-sleep scheduling for a sender-receiver pair in \gls{siot} as a \gls{mdp}. In this \gls{mdp}, actions correspond to deciding whether the device enters deep-sleep mode and, if so, for how long. Whenever the \gls{siot} device is awake, it has to determine its next deep-sleep duration to minimize a weighted sum of costs derived from the \gls{got} metric and energy costs.
    \item In the formulated \gls{mdp}, we model the device's current \emph{belief}, which serves as the basis for the device's decision about its next deep-sleep duration. This belief consists of all the information, which the device learned previously about the channel conditions and the observed process.
    \item We propose a novel \sysname algorithm for optimizing these decisions on deep-sleep durations. \sysname uses learned knowledge about the observed process and the channels to carry out simulations for different deep-sleep durations. Based on the expected time-averaged cost values obtained through these simulations, \sysname decides on the next deep-sleep duration.
    \item Through extensive numerical simulations, we analyze \sysname's behavior under varying conditions and show its superior performance compared to baselines, such as threshold-based solutions and Q-learning.
    \item We validate the feasibility and effectiveness of our approach through a prototype implementation on real \gls{siot} hardware, demonstrating its practical applicability.
\end{itemize}

The paper is organized as follows: Section \ref{sec:sota} reviews related work. Section \ref{sec:system_model} introduces the system model, and Section \ref{sec:problemformulation} formulates the optimization problem. Our approach, \sysname, is described in Section \ref{sec:approach}.
In Section \ref{sec:simulations} we present numerical and experimental results, respectively, with implementation details in Section \ref{sec:implementation}.
\rev{In Section \ref{sec:discussion}, we discuss limitations of the used model and our proposed method.}
Section \ref{sec:conclusions} concludes the paper.

\section{RELATED WORK}
\label{sec:sota}

This section reviews related work on deep-sleep scheduling for minimizing energy consumption and for content-aware, age-based metrics in the \gls{siot}. We categorize and summarize existing works and conclude this section by describing how our approach differs from the state of the art.

Table \ref{tab:sota} compares the studies discussed in this section across the above-mentioned criteria.
Broadly, the literature falls into two groups. 
First, studies that optimize deep-sleep scheduling but do not use a content-aware age-based metric \cite{Wang2024_2, Badia2025, Cao2023, Gong2022}.
Second, studies that employ the \gls{aoii}, a content-aware age-based metric, but do not address deep-sleep scheduling \cite{Dongare2024, Chen2021, Chen2024, Kam2020, deSombre2025b, Hong2024, Kriouile2022, Maatouk2022, deSombre2023b, Yan2024}.
The authors of \cite{Gursoy2024} offer first insights into the combination of both lines of research.

We begin with the first group. The optimization of deep-sleep intervals has primarily been studied in the context of \gls{aoi} \cite{Wang2024_2, Badia2025, Cao2023, Gong2022}, where only the authors of \cite{Badia2025} explicitly consider the \gls{siot} and its distinctive channel conditions in their model. 

Specifically, in \cite{Wang2024_2}, deep-sleep schedules in the \gls{iiot} are optimized using Proximal Policy Optimization.
The authors of \cite{Cao2023} use an \gls{mdp} to derive an optimal cyclic deep-sleep schedule, offering a closed-form solution for the \gls{aoi}-energy tradeoff.
In \cite{Gong2022}, deep-sleep intervals are scheduled for unreliable channels using \gls{mdp}-based methods, providing a two-threshold optimal policy with a closed-form expression for the AoI-energy tradeoff.
The authors of \cite{Badia2025} schedule transmissions in the \gls{siot} to minimize the \gls{aoi} over a finite time horizon.

While the above-cited papers make essential contributions to optimizing deep-sleep schedules for age-based metrics, these approaches focus exclusively on minimizing information age and do not consider a content-aware metric, which can lead to unnecessary transmissions.

In recent years, a new line of research has emerged around the \gls{aoii} (also referred to as \gls{aos}), introduced in \cite{Maatouk2020}.
Unlike \gls{aoi}, the \gls{aoii} accounts for the age of information and also for the similarity between the sender's and receiver's information.
The \gls{aoii} has been studied in various scenarios.

In \cite{Dongare2024}, we minimize the \gls{aoii} in energy-harvesting \gls{iot} systems by learning transmission schedules through a \gls{fdpg} method.
Similarly, in \cite{Chen2021} and \cite{Chen2024}, the authors model unreliable channels or channels with random delay and develop threshold-based scheduling strategies using constrained or standard Markov decision processes.
These works prove the optimality of simple policies under certain system conditions.

In \cite{Kam2020}, the authors minimize the \gls{aoii} for a binary Markov source, using dynamic programming to derive optimal sampling policies.
The authors of \cite{Kriouile2022, Maatouk2022} investigate the trade-off between energy consumption and \gls{aoii}. 
We analyze the same trade-off in our previous work \cite{deSombre2023b}.
The authors of \cite{Hong2024} and \cite{Yan2024} consider transmission scheduling to optimize the \gls{aoii} in the context of the \gls{siot}.

Beyond the transmitter side, recent works also explore optimization at the receiver or network level.
All the authors of these works provide important insights into minimizing the \gls{aoii}, but do not optimize deep-sleep intervals.

Only in \cite{Gursoy2024}, deep-sleep intervals are optimized to minimize both \gls{aoii} and energy consumption using threshold-based methods. 
However, the authors assume a fixed cycle length, which is suboptimal. Furthermore, the work focuses on minimizing energy consumption subject to an \gls{aoii} constraint, rather than considering a more flexible weighted sum of \valname degradation and energy consumption.

The concept of the \gls{aoii} was further generalized in \cite{Li2024}, where Li et al. introduced the \gls{got} metric. 
This metric allows to use application-specific weights to scale costs of \valname degradation depending on a basic \valname degradation metric like either the \gls{aoi} or the \gls{aoii}, the current state of the observed process and the information at the receiver.
By using the \gls{got} metric for deep-sleep scheduling, we extend and generalize the discussed line of research on age-based metrics in the \gls{iot}.

\section{SYSTEM MODEL}
\label{sec:system_model}

\begin{figure}[t]
    \centering
    \includegraphics[width=\columnwidth, trim=0cm 2cm 0cm 0, clip]{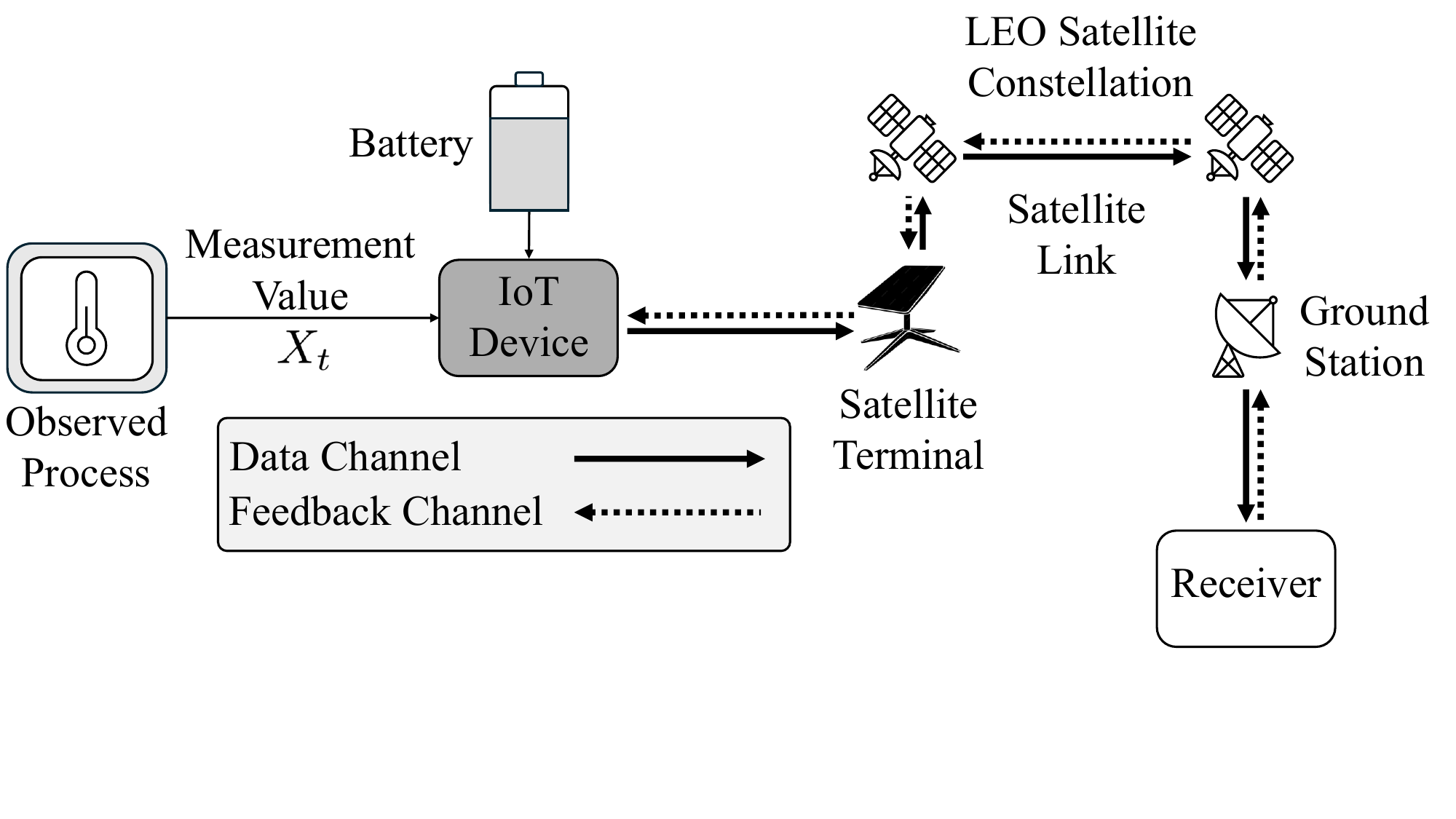}
    \caption{System Model}
    \label{fig:systemmodel}
\end{figure}

Fig. \ref{fig:systemmodel} illustrates the main components of the considered system. 
It consists of a battery-powered \gls{iot} device that transmits environmental measurements via a multi-hop connection to a receiving server (called \textit{receiver}).
The link between the \gls{iot} device and the receiver comprises a wireless connection to a satellite terminal, the wireless link from the satellite terminal to a satellite of the satellite constellation, inter-satellite-links within the constellation, the connection from a satellite to a ground station, and the connection between ground station and receiver. 

\begin{figure}[t]
    \centering
    \includegraphics[width=\columnwidth]{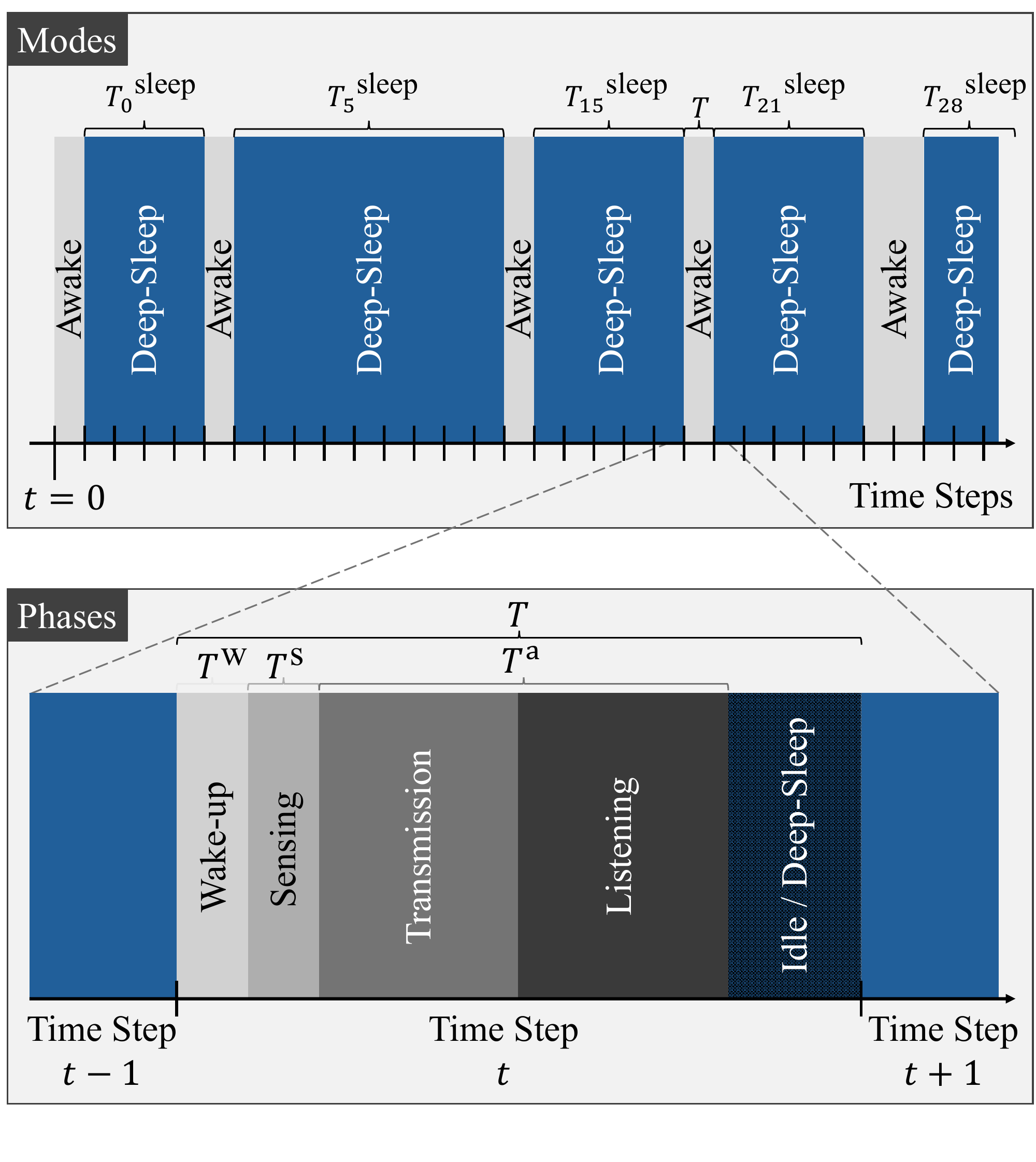}
    \caption{Overview of the device’s operation schedule. Top: Alternation between deep-sleep and awake mode across multiple time steps. Bottom: Detailed view of a single time step.
    }
    \label{fig:timesteps}
    \vspace{-0.3cm}
\end{figure}

We consider an infinite time horizon, divided into discrete time steps $t=0,1,\dots$, each of the same duration $T$.
As indicated in Fig. \ref{fig:timesteps}, the \gls{iot} device is in one of two possible modes, either awake or in deep-sleep. 
If it is awake, it can use its sensors, power its antenna, and decide to switch to deep-sleep mode. 
If the device decides to sleep after time step $t$, it specifies a number $N_t^\mathrm{sleep}$ of deep-sleep time steps and sets a timer.  
Then, after $N_t^\mathrm{sleep}$ time steps, the device wakes up automatically.

In a single time step in which the device is in awake mode, it progresses through the following phases in a fixed order as indicated in Fig. \ref{fig:timesteps}. All phases are assumed to have fixed duration, except for the Listening-Phase whose length depends on when feedback is received:
\begin{itemize}
  \item \textit{Wake-up Phase}: The IoT device powers up its core components and prepares for operation.
  \item \textit{Sensing-Phase}: The device uses its sensor to sample the observed process and stores the new measurement in its buffer.
  \item \textit{Transmission-Phase}: The device uses its antenna to transmit the new measurement.
  \item \textit{Listening-Phase}: The device waits for feedback from the receiver.
  In case no ACK is received in this phase, the device assumes the transmission failed.
  \item \textit{Idle-Phase}: After feedback is received, the device turns off its antenna and waits until the beginning of the next time step. If the device has decided to sleep in the next time step, this phase is skipped and the device transitions directly to deep-sleep mode. This transition is assumed to be practically instantaneous, so no separate fall-asleep phase is modeled.
\end{itemize}
In deep-sleep mode, there are no distinct phases.

The \gls{iot} device observes an environmental process such as electric current, temperature, acceleration, humidity or light intensity.
This process is shown on the left in Fig. \ref{fig:systemmodel}. 
The dynamics of this process are captured by a Markov Chain $\mathcal{X}$, which is assumed to be stationary. 
The process state in time step $t$ is denoted by $X_t$.

We represent the observed process by a discrete state space. 
Therefore, physical values are quantized into one of the finitely many states in $\mathcal{X}$, and $X_t$ denotes the resulting quantized process state.
The evolution of $X_t$ follows the transition probabilities given by the matrix $P^\mathrm{proc}$, which is defined as:
\begin{equation}
    \label{eq:process_evolution}
    P^\mathrm{proc}(X, X') := \mathbb{P}(X_{t+1} = X' \mid X_t = X).
\end{equation}

To measure the observed process, the \gls{iot} device uses its sensors and stores the latest measurement in its buffer. 
The measurement in the device's buffer in time step $t$ are denoted by $X_t^\mathrm{Tx}$.

After sensing, the \gls{iot} device transmits the new measurement over a multi-hop connection, outlined as solid arrows in Fig. \ref{fig:systemmodel}, to the receiver.
We call this connection the \emph{data channel}.
In case the age-based metric $F$ used as input for the \gls{got} metric is the \gls{aoi}, we assume that the \gls{iot} device transmits every new measurement. 
In case $F$ is the \gls{aoii}, the device transmits a new measurement only if it differs from the information currently available at the receiver.

As shown in Fig. \ref{fig:systemmodel}, the transmission starts with the device forwarding its buffered measurement to a satellite terminal. 
From this satellite terminal, the measurement is sent to a satellite of a \gls{leo} satellite constellation using Ka-Band frequencies.
The measurement is then directly transmitted to a terrestrial ground station over Ku-Band or indirectly via optical inter-satellite links through another satellite.
Finally, the ground station delivers the measurement via a fiber-optic connection to the receiving server.
The most recently received measurement at the receiver in time step $t$ is denoted by $X_t^\mathrm{Rx}$.

In case the data reaches the receiver successfully, an acknowledgment signal is transmitted via the feedback channel to the \gls{iot} device.
In Fig. \ref{fig:systemmodel}, the feedback channel is outlined as dotted arrows.
The acknowledgment is routed back to the \gls{iot} device by retracing the same transmission steps as the measurement data, but in reverse order.
Note that while the transmission steps are mirrored, the physical route may differ.

We model the data channel from the \gls{iot} device to the receiver as a Markov channel (cf. \cite{Wang1995}).
A Markov channel models the channel as a packet erasure channel with a packet erasure rate and a delay changing according to a Markov chain.
Every channel state $C_\mathrm{data} \in \mathcal{C}^\mathrm{data}$ is a tuple of two values $C_\mathrm{data} = (C_\mathrm{data}^ \mathrm{delay}, C_\mathrm{data}^\mathrm{erasure})$, where the first value, $C_\mathrm{data}^\mathrm{delay}$, models the channel's delay in seconds.
The second value, $C_\mathrm{data}^\mathrm{erasure}$, models the channel's erasure rate.
Whenever the \gls{iot} device attempts to transmit a measurement, the data reaches with a probability of $1-C_\mathrm{data}^\mathrm{erasure}$, and is erased with a probability of $C_\mathrm{data}^\mathrm{erasure}$. 

We model the state of the feedback channel separately from that of the data channel as $C_\mathrm{feedback} = (C_\mathrm{feedback}^ \mathrm{delay}, C_\mathrm{feedback}^\mathrm{erasure})$.

We model their dynamics using the Markov chain $\mathcal{C}$ with states $C = (C_\mathrm{data}, C_\mathrm{feedback})$.
The transition probabilities for this Markov chain are given by the matrix $P^\mathrm{channel}$ defined by:
\begin{equation}
    \label{eq:channel_evolution}
    P^\mathrm{channel}(C, C') := \mathbb{P}(C_{t+1} = C'|C_t=C).
\end{equation}
Note that although the states of the two channels can correlate, they are not the same.

The device operates on battery power, and each action draws electrical power at a different level.
Waking up from deep-sleep has a time duration $T^\mathrm{w}$ (in seconds) and draws power $P^\mathrm{w}$.
Sensing has a time duration $T^\mathrm{s}$ (in seconds) and draws power $P^\mathrm{s}$.
Transmitting and receiving feedback requires activating the antenna and draws power $P^\mathrm{a}$.
Deep-sleep mode incurs a minimal baseline power draw $P^\mathrm{d}$, and idling incurs a power draw $P^\mathrm{i}$.

The device aims to balance its energy consumption and the \valname degradation at the receiver. Here, the latter is measured using the \gls{got} metric \cite{Li2024}. The \gls{got} metric builds on classical age-based metrics like either the \gls{aoi} or the \gls{aoii}. We first define these age-based metrics before providing a definition for the \gls{got} metric. 

With $\mathrm{AoI}_0 = 0 $, the \gls{aoi} at time step $t$ is recursively defined as
\begin{align}
    \mathrm{AoI}_{t+1} = \begin{cases}
        0, & \text{if successfully transmitted}, \\
        \mathrm{AoI}_t + 1, & \text{otherwise}.
    \end{cases}
\label{eq:aoi_update}
\end{align}

The \gls{aoii} combines both freshness and correctness, and it is defined recursively as
\begin{align}
    \mathrm{AoII}_{t+1} = \begin{cases}
        \mathrm{AoII}_t + 1, & \text{if } X_t \neq X_t^\mathrm{Rx}, \quad \quad \quad \quad \quad \quad\\
        0, & \text{otherwise},
    \end{cases}
\label{eq:aoii_update}
\end{align}
with $\mathrm{AoII}_0 = 0 $. 
Assuming that after $M$ time steps, data is outdated in the sense that the \valname will not further decrease, we use a cap $M \in \mathbb{N}$ for age-based metrics, meaning that instead of increasing indefinitely, these metrics grow until they reach the value $M$ and stay constant at $M$ afterwards.

The \gls{got} metric is defined as a mapping
\begin{equation}
    \mathrm{GoT}: \mathcal{X} \times \mathcal{X} \times \{0, \dots, M\} \rightarrow \mathbb{R},
\end{equation}
where $\mathcal{X}$ denotes the space of possible process states, and $M$ is the maximum value of the considered age-based metric $F$.
Notably, widely-used metrics such as \gls{mse}, \gls{aoi}, and \gls{aoii} can be represented as special cases of the \gls{got}. 

To illustrate the advantages of the \gls{got}, we provide a detailed example:
\begin{example}
    In safety-critical systems, certain mismatches are more dangerous than others. Consider the case where $X_t \in \{\texttt{Safe}, \, \texttt{Warning}, \, \texttt{Emergency}\}$, and the receiver state $X_t^\mathrm{Rx} \in \{\texttt{Safe}, \, \texttt{Warning}, \, \texttt{Emergency}\}$.
    Then we can define the function $\mathrm{GoT}$ as follows:
    \begin{align}
        \mathrm{GoT}(X_t, X_t^\mathrm{Rx}, \mathrm{AoI}_t) = &\alpha \cdot \mathbb{1}_{\{X_t = \texttt{Emergency} \land X_t^\mathrm{Rx} = \texttt{Safe}\}}  \nonumber \\
        &+ \beta \mathrm{AoI}_t, \nonumber
    \end{align}
    where $\alpha \gg \beta$ assigns a high penalty if the \gls{iot} device detects an emergency, but the receiver is unaware, which is particularly hazardous.
\end{example}

\rev{In practice, constructing the \gls{got} starts from the application objective and the consequences of incorrect information at the receiver.
For each pair $(X_t, X^{\mathrm{Rx}}_t)$, a penalty value is specified that reflects the criticality of this mismatch for the application. 
The dependence of the \gls{got} metric on the age-based metric $F$ determines how the cost increases as outdated information persists.
Example 1 follows this construction. The pair $X_t = \texttt{Emergency}$ and $X^{\mathrm{Rx}}_t = \texttt{Safe}$ is assigned a substantially larger penalty than other pairs, while the age-dependent term captures the additional cost caused by increasingly outdated information at the receiver.
We refer the reader to \cite{Li2024} for a more general step-by-step \gls{got} construction methodology and further examples.}

\section{PROBLEM FORMULATION}
\label{sec:problemformulation}

The goal of the \gls{iot} device is to find a tradeoff between its energy consumption and \valname degradation at the receiver.  
To this end, the device decides whether to enter deep-sleep mode and, if so, for how long. 
This decision-making process is based on the available information that the device continuously collects and utilizes.

We formulate this decision-making problem as an \gls{mdp} $\mathcal{M} = (\mathcal{S}, \mathcal{A}, c, P)$ consisting of a set $\mathcal{S}$ of states, a set $\mathcal{A}$ of actions, a cost function $c: \mathcal{S} \times \mathcal{A} \rightarrow \mathbb{R}$, and transition probabilities $P: \mathcal{S} \times \mathcal{A} \times \mathcal{S} \rightarrow [0,1]$.
In this section, we define the components of this \gls{mdp} and derive the corresponding optimization problem.

To define the states of $\mathcal{M}$, we first define the notion of a \emph{belief} at the \gls{iot} device.
In time step $t$, the belief $B_t \in \mathcal{B}$ is a tuple consisting of
\begin{itemize}
    \item the state $X_t^\mathrm{Tx} \in \mathcal{X}$ in the device's buffer, representing the most recently sensed measurement from the observed process,
    \item estimated transition probabilities $P_t^\mathrm{proc, est}: \mathcal{X} \times \mathcal{X} \rightarrow [0,1]^{}$ of the observed process,
    \item the current belief distribution $D_t^{X}: \mathcal{X} \rightarrow [0,1]$ over the states of the observed process,
    \item the belief distribution $D_t^{X^\mathrm{Rx}}: \mathcal{X} \rightarrow [0,1]$ over the states of the observed process most recently received at the receiver,
    \item the belief distribution of the age-based metric used as input for the \gls{got}, which is either
    \begin{itemize}
        \item the belief distribution $D_t^{\mathrm{AoI}^\mathrm{Rx}}: (0,...,M) \rightarrow [0,1]$ over the \gls{aoi} at the receiver, or
        \item the tensor $T_t^\mathrm{AoII}: \mathcal{X} \times \mathcal{X} \times (0,...,M) \rightarrow [0,1]$, containing the device's knowledge about the current \gls{aoii} depending on $X$ and $X^\mathrm{Rx}$,
    \end{itemize}
    \item the sleep duration $N_t^\mathrm{sleep} \in \mathbb{N}_0$ indicating the number of time steps the device will stay in deep-sleep before waking up, and
    \item the last time step $t_t^\mathrm{awake}$ during which the device was awake.
\end{itemize}

Using this definition of a belief, we define a state $S=(C, X, X^\mathrm{Tx}, X^\mathrm{Rx}, \mathrm{AoI}^\mathrm{Tx}, \mathrm{AoI}^\mathrm{Rx}, \mathrm{AoII}, B) \in \mathcal{S}$. It consists of:
\vspace{-0.2cm}
\begin{itemize}
    \item a channel state $C \in \mathcal{C}$,
    \item a state $X \in \mathcal{X}$ of the observed process,
    \item a state $X^\mathrm{Rx} \in \mathcal{X}$ most recently received at the receiver,
    \item an age of information $\mathrm{AoI}^\mathrm{Rx} \in (0,...,M)$ at the receiver,
    \item an age of incorrect information $\mathrm{AoII} \in (0,...,M)$, and
    \item a belief $B \in \mathcal{B}$ at the \gls{iot} device as described above.
\end{itemize}
The state of the \gls{mdp} at time step $t$, denoted by $S_t$, consists of the same components, each indexed by $t$.

The set $\mathcal{A} = \{0,1,\dots,a_\mathrm{max}\}$ of actions consists of all possible deep-sleep intervals ($a_t = 0$ denotes staying awake). 
$a_t$ specifies the duration of the deep-sleep after time step $t$.
To determine action $a_t \in \mathcal{A}$ in time step $t$, the device can only access the information in $B_t$. 
A strategy $\pi$ is called a \textit{belief-based strategy}, if it is a mapping $\pi: \mathcal{B} \rightarrow \mathcal{A}$.

The transition probabilities $P$ of $\mathcal{M}$ are given by the dynamics described in Section \ref{sec:system_model} and can be summarized as follows:
The evolution of $X_t$ and $C_t$, given by (\ref{eq:process_evolution}) and (\ref{eq:channel_evolution}), is assumed to be stochastically independent.
$X^\mathrm{Rx}$ is updated in case of a successful transmission as
\begin{equation}
    X_{t+1}^\mathrm{Rx} = \begin{cases}
        X_{t+1}^\mathrm{Tx} &\text{ if successfully transmitted}, \\
        X_{t}^\mathrm{Rx} &\text{ otherwise}.
    \end{cases}
\end{equation}
Whether a transmission attempt is successful depends on the current data channel quality $C_{\mathrm{data}, t}^\mathrm{erasure}$ as described in Section \ref{sec:system_model}.
$\mathrm{AoI}_t^\mathrm{Rx}$ and $\mathrm{AoII}_t$ are updated according to (\ref{eq:aoi_update}) and (\ref{eq:aoii_update}).
The transition from $B_t$ to $B_{t+1}$ is outlined in Alg. \ref{alg:update}, which is described in detail in Sec. \ref{sec:approach}.

Energy costs are defined as:
\begin{equation}
    \label{eq:energy_costs}
    c_\mathrm{e}(S_t) = T_t^\mathrm{w} \cdot P^\mathrm{w} + T_t^\mathrm{s} \cdot P^\mathrm{s} + T_t^\mathrm{a} \cdot P^\mathrm{a} + T_t^\mathrm{i} \cdot P^\mathrm{i} + T_t^\mathrm{d} \cdot P^\mathrm{d},
\end{equation}
where $T_t^\mathrm{w}$ is the time spent waking up in time step $t$. 
If the device was already awake in the previous time step $t-1$, then $T_t^\mathrm{w} = 0$.
$T_t^\mathrm{s}$, $T_t^\mathrm{a}$, $T_t^\mathrm{i}$ and $T_t^\mathrm{d}$ represent the time the device spends using sensors, powering the antenna, idling and remaining in deep-sleep, respectively.
The sum of all durations per time step is constant: $T_t^\mathrm{w} + T_t^\mathrm{s} + T_t^\mathrm{a} + T_t^\mathrm{d} = T$.

Without new transmissions, the \valname at the receiver degrades. The cost incurred by this degradation is given by:
\begin{align}
    \label{eq:got_costs}
    c_\mathrm{\valindex}(S_t) &= c_\mathrm{\valindex}((X, X^\mathrm{Tx}, X^\mathrm{Rx}, \mathrm{AoI}^\mathrm{Tx}, \mathrm{AoI}^\mathrm{Rx}, \mathrm{AoII}^\mathrm{Rx})) \nonumber \\
    &= \mathrm{GoT}(X, X^\mathrm{Rx}, F),
\end{align}
where $F$ is either $\mathrm{AoI}^\mathrm{Rx}$ or $\mathrm{AoII}^\mathrm{Rx}$.

Together, we define the cost function $c: \mathcal{S} \rightarrow \mathbb{R}$ of $\mathcal{M}$ as a weighted sum of energy consumption cost and \valname degradation cost:
\begin{equation}
    \label{eq:cost}
    c(S_t) = w_\mathrm{e} \cdot c_\mathrm{e}(S_t) + w_\mathrm{\valindex} \cdot c_\mathrm{\valindex}(S_{t}).
\end{equation}
Here, $w_\mathrm{e}, w_\mathrm{\valindex} \in \mathbb{R}$ are weights that define the relative importance of energy consumption and \valname in the device's optimization objective.

Using the defined terms, the average cost of a strategy $\pi$ can be expressed as
\begin{equation}
    c^\pi_\mathrm{avg} = \lim_{t_\mathrm{final} \rightarrow \infty} \left[ \frac{1}{t_\mathrm{final}} \sum_{t=0}^{t_\mathrm{final} - 1} c(S^\pi_{t}) \right],
\end{equation}
where the states $S^\pi_t$ are obtained using $\mathcal{M}$ and $\pi$.
Accordingly, we define 
\begin{equation}
    c^\pi_\mathrm{\valindex} = \lim_{t_\mathrm{final} \rightarrow \infty} \left[ \frac{1}{t_\mathrm{final}} \sum_{t=0}^{t_\mathrm{final} - 1} c_\mathrm{\valindex}(S^\pi_{t}) \right]
\end{equation}
and
\begin{equation}
    c^\pi_\mathrm{e} = \lim_{t_\mathrm{final} \rightarrow \infty} \left[ \frac{1}{t_\mathrm{final}} \sum_{t=0}^{t_\mathrm{final} - 1} c_\mathrm{e}(S^\pi_{t}) \right].
\end{equation}
With $c^\pi_\mathrm{avg}$, the optimization problem can be written as:
\begin{equation}
    \min_{\pi \in \Pi} \mathbb{E}  [c^\pi_\mathrm{avg}].
\end{equation}
Consequently, we refer to the optimized quantity as \textit{expected time-averaged cost}.

\section{Probabilistic Simulation-based Optimization}
\label{sec:approach}

\begin{figure}[t]
    \centering
    \includegraphics[width=\columnwidth]{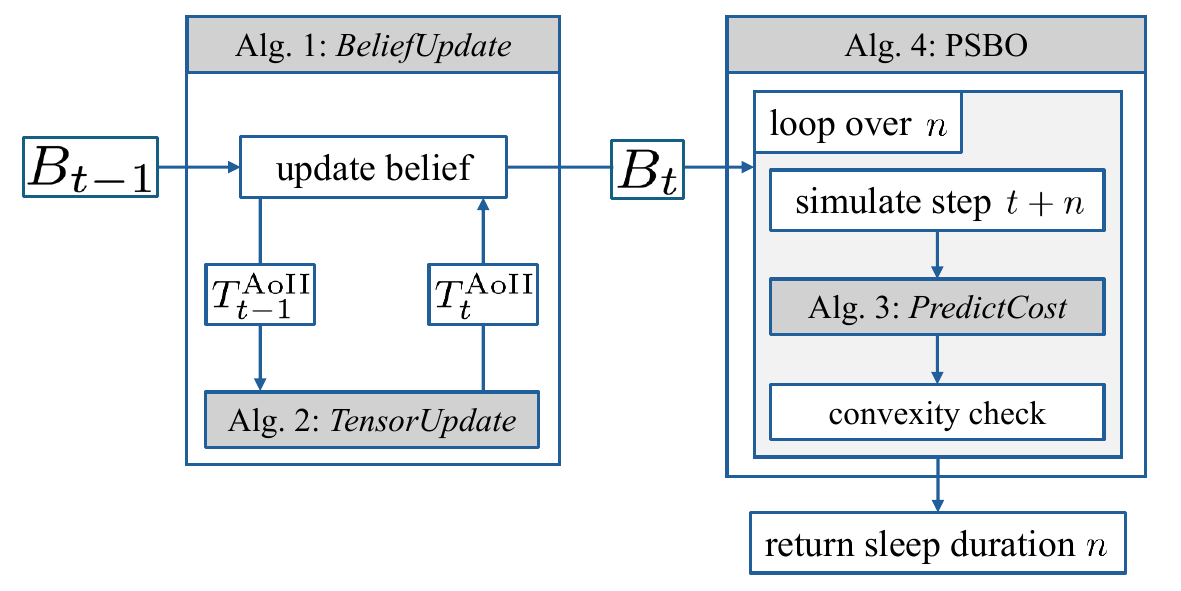}
    \caption{Overview of the interaction between Alg. 1 to 4.}
    \label{fig:flowchart}
\end{figure}

In this section, we introduce \sysname, our approach for finding the optimal tradeoff between \valname degradation at the receiver and the energy consumption of the \gls{iot} device.
To the best of our knowledge, we are the first to apply this type of algorithm based on probabilistic simulations to treat the tradeoff of energy consumption and age-based metrics for \valname degradation.

For an overview over our solution approach, we provide a flowchart including the involved algorithms in Fig. \ref{fig:flowchart}.
Following the procedure over the course of a single time step $t$, we describe these algorithms in Sec. \ref{sec:belief_update} to Sec. \ref{sec:psbo}.

\rev{At a high level, \sysname operates in three stages during each decision epoch.} 

\rev{First, the device updates its belief about the observed process, the receiver state, and the age-based metric from the latest observation and feedback. 
This update is performed by Alg. \ref{alg:update}. 
If the age-based metric is AoI, the corresponding belief is updated directly.
If the age-based metric is \gls{aoii}, Alg. \ref{alg:update} uses Alg. \ref{alg:tensor} as a sub-procedure to update the corresponding belief distribution based on the current state of the observed process and the current state at the receiver.} 

\rev{Second, starting from the updated belief, Alg. \ref{alg:approach} evaluates candidate deep-sleep durations by propagating the belief through a distribution-based simulation. 
For each candidate duration, it uses Alg. \ref{alg:cost} to compute the expected one-step \gls{got} cost and the resulting expected time-averaged cost after waking up.}

\rev{Third, PSBO repeats the evaluation described above for increasing deep-sleep durations and performs a convexity check after each iteration to finally return the optimal deep-sleep duration.}

\rev{The individual algorithms are presented in the following order. 
In Sec. \ref{sec:belief_update}, we describe Alg. \ref{alg:update} and its \gls{aoii}-specific sub-procedure Alg. \ref{alg:tensor}.
In Sec. \ref{sec:cost_prediction}, we explain Alg. \ref{alg:cost}. 
Finally, in Sec. \ref{sec:psbo}, we describe Alg. \ref{alg:approach}, which combines these components to search over candidate deep-sleep durations.
}

\subsection{Belief Update}
\label{sec:belief_update}
\begin{algorithm}[t]
\caption{\textit{BeliefUpdate}}
\label{alg:update}
\DontPrintSemicolon
\SetKwInOut{Input}{Input}
\SetKwInOut{Output}{Output}
\begin{fivept}
\Input{
    Belief $B_{t-1}$
}
\Output{Updated belief $B_t$}

\uIf{$t_{t-1}^\mathrm{awake} + N_{t-1}^\mathrm{sleep} > t$}{
    $X_t^\mathrm{Tx} \gets X_{t-1}^\mathrm{Tx}$\;
    $D_t^{X} \gets D_{t-1}^{X} \cdot P_{t-1}^\mathrm{proc, est}$\;
}
\Else{
    $X_t^\mathrm{Tx} \gets X_t$\;
    $P_t^\mathrm{proc, est} \gets t \cdot P_{t-1}^\mathrm{proc, est}$\;
    \uIf{$X_{t-1}^\mathrm{Tx} = X_t$}{
        $P_t^\mathrm{proc, est}(X_{t-1}^\mathrm{Tx}, X_{t-1}^\mathrm{Tx}) \gets P_{t-1}^\mathrm{proc, est}(X_{t-1}^\mathrm{Tx}, X_{t-1}^\mathrm{Tx}) + N_{t-1}^\mathrm{sleep} + 1$\;
    }
    \Else{
        $P_t^\mathrm{proc, est}(X_{t-1}^\mathrm{Tx}, X_{t-1}^\mathrm{Tx}) \gets P_{t-1}^\mathrm{proc, est}(X_{t-1}^\mathrm{Tx}, X_{t-1}^\mathrm{Tx}) + \lfloor N_{t-1}^\mathrm{sleep} / 2 \rfloor$\;
        $P_t^\mathrm{proc, est}(X_{t-1}^\mathrm{Tx}, X_t^\mathrm{Tx}) \gets P_{t-1}^\mathrm{proc, est}(X_{t-1}^\mathrm{Tx}, X_t^\mathrm{Tx}) + 1$\;
        $P_t^\mathrm{proc, est}(X_t^\mathrm{Tx}, X_t^\mathrm{Tx}) \gets P_{t-1}^\mathrm{proc, est}(X_t^\mathrm{Tx}, X_t^\mathrm{Tx}) + N_{t-1}^\mathrm{sleep} - \lfloor N_{t-1}^\mathrm{sleep} / 2 \rfloor$\;
    }
    Normalize $P_t^\mathrm{proc, est}$ row-wise\;
    $D_t^{X} \gets (0, \dotsc, 0)$\;
    $D_t^{X}(X_t^\mathrm{Tx}) \gets 1$\;
    $D_t^{X^\mathrm{Rx}} \gets (0, \dotsc, 0)$\;
    $D_t^{X^\mathrm{Rx}}(X_t^\mathrm{Rx}) \gets 1$\;
    $t_{t-1}^\mathrm{awake} \gets t$\;
}
\uIf{$F = \mathrm{AoI}$}{
    $D_t^{\mathrm{AoI}^\mathrm{Rx}} \gets (0, \dotsc, 0)$\;
    $D_t^{\mathrm{AoI}^\mathrm{Rx}}(\mathrm{AoI}_t^\mathrm{Rx}) \gets 1$\;
}
\ElseIf{$F = \mathrm{AoII}$}{
    $T_t^{\mathrm{AoII}} \gets \mathit{TensorUpdate}(B_{t-1}, D_t^{X})$\;
}
\Return $B_t$\;
\end{fivept}
\end{algorithm}

\begin{algorithm}[t]
\caption{\textit{TensorUpdate}}
\label{alg:tensor}
\DontPrintSemicolon
\SetKwInOut{Input}{Input}
\SetKwInOut{Output}{Output}
\begin{fivept}
\Input{Belief $B_{t-1}$, Updated Distribution $D_t^{X}$}
\Output{Updated Tensor $T_t$}
$T_t^{\mathrm{AoII}} \gets 0$\;
\ForEach{$X \in \mathcal{X}$}{
    \If{$D_{t-1}^{X}(X) > 0$}{
        \ForEach{$X' \in \mathcal{X}$}{
            $R \gets (0,\dotsc,0)$\;
            \For{$k = 0$ \KwTo $M-1$}{
                $R(k+1) \gets T_{t-1}^{\mathrm{AoII}}(X, X', k)$\;
            }
            $R(M) \gets R(M) + T_{t-1}^{\mathrm{AoII}}(X, X', M)$\;
            \ForEach{$X'' \in \mathcal{X}$}{
                \If{$D_t^{X}(X'') > 0$}{
                    \uIf{$X'' \neq X'$}{
                        \uIf{$D_t^{X}(X'') = 1$}{
                            \For{$\Delta = 0$ \KwTo $M$}{
                                $T_t^{\mathrm{AoII}}(X'', X', \Delta) \mathrel{+}= D_{t-1}^{X}(X) \cdot R(\Delta)$\;
                            }
                        }
                        \Else{
                            \For{$\Delta = 0$ \KwTo $M$}{
                                $T_t^{\mathrm{AoII}}(X'', X', \Delta) \mathrel{+}= \frac{D_{t-1}^{X}(X)}{D_t^{X}(X'')} \cdot P_{t-1}^\mathrm{proc, est}(X, X'') \cdot R(\Delta)$\;
                            }
                        }
                    }
                    \Else{
                        $T_t^{\mathrm{AoII}}(X'', X', 0) \gets 1$\;
                    }
                }
            }
        }
    }
}
\Return $T^\mathrm{AoII}_t$\;
\end{fivept}
\end{algorithm}

\subsubsection{\rev{Intuition behind Algorithm \ref{alg:update} and Algorithm \ref{alg:tensor}}}
\rev{At a high level, Alg. \ref{alg:update} has two cases. 
If the device is still in deep-sleep, only the belief about the observed process is propagated one step forward. 
If the device is awake, the new measurement is stored, the estimated process transition probabilities are updated, and the belief about the state at the receiver is reset based on the currently available ACK information. 
The final step in Alg. 1 depends on the chosen age-based metric. 
For the \gls{aoi}, the age distribution is reset directly. 
For the \gls{aoii}, Alg. \ref{alg:tensor} updates the tensor containing the \gls{aoii} distributions.}

\subsubsection{\rev{Description of Algorithm \ref{alg:update}}}

Alg. \ref{alg:update} describes the procedure for updating the \gls{iot} device's belief from $B_{t-1}$ to $B_{t}$ at time step $t$.
Here, we update every component of the belief as listed in Section \ref{sec:system_model}.

Lines 1–3 check whether the device was in deep-sleep mode during time step $t$.
If it was, the buffered measurement $X_{t}^{\mathrm{Tx}}$ remains unchanged compared to $X_{t-1}^{\mathrm{Tx}}$, and the device updates its belief distribution $D_{t}^{X}$ about the environmental process by multiplying with the estimated transition probabilities $P_{t-1}^{\mathrm{proc,est}}$.
This way, we simulate one step of the observed process using the knowledge gathered over time in the form of the estimated transition probabilities.

Lines 4–17 correspond to the case when the device is awake and performs sensing: the new measurement is stored in the buffer (line 5), and the estimated transition probabilities $P_{t-1}^{\mathrm{proc,est}}$ are updated based on the observed state transitions during deep-sleep mode (lines 6–13).
This update assumes that at most one state transition occurred during the sleep period.
This assumption does not necessarily hold true.

However, for fast changing processes, our solution tends to choose shorter deep-sleep intervals, which reduced potential errors.
Alternatively, it is possible to initially spend energy to stay awake for a certain period of time to estimate the transition probabilities of the observed process without using the additional above-mentioned assumption.

In lines 14-17, the belief distributions $D_{t}^{X}$ and $D_{t}^{X^\mathrm{Rx}}$ are reset, by setting the probability to the current states to $1$. 
Line 18 marks the current time as the last awake time step.
If the age-based metric $F$ used as the input for the \gls{got} is the \gls{aoi}, we update the \gls{aoi} distribution $D_{t-1}^{\mathrm{AoI}^\mathrm{Rx}}$ according to Eq. \ref{eq:aoi_update} (line 20 and 21).
Both in line 17 and 21, the device only assumes that the transmission was successful in case it received an ACK over the feedback channel.

If $F$ is the \gls{aoii}, line 23 of Alg. \ref{alg:update} updates the tensor $T_{t-1}^{\mathrm{AoII}}$, which captures the distributions over the \gls{aoii}.
This tensor update is performed using Alg. \ref{alg:tensor}.
The updated belief $B_{t}$ is returned in line 24.
To update its belief, the \gls{iot} device performs this update $N^\mathrm{sleep}_{t-1}$ times after waking up.
\rev{\subsubsection{Time complexity of Algorithm \ref{alg:update}}
For $F=\mathrm{AoI}$, the time complexity of Algorithm \ref{alg:update} is in $\mathcal{O}(|\mathcal{X}|^2 + M)$, where the additional $M$ is due to line 20.
For $F=\mathrm{AoII}$, the time complexity of Algorithm \ref{alg:update} is dominated by the time complexity of Algorithm \ref{alg:tensor} in line 23 and therefore has time complexity in $\mathcal{O}(|\mathcal{X}|^3 M)$.}
\subsubsection{\rev{Description of Algorithm \ref{alg:tensor}}}
Next, we explain Alg. \ref{alg:tensor}, which is used in line 23 of Alg. \ref{alg:update} and updates the device's belief distribution $T_t^{\mathrm{AoII}}$ for the \gls{aoii}.
To update every entry of the tensor $T_t^{\mathrm{AoII}}$, we must consider every possible transition in the process state.

We iterate over all possible previous process states $X$, all possible states $X'$ at the receiver and all possible current states $X''$. 
Lastly, we have to iterate over all possible values $\Delta$ of the \gls{aoii}.

In the following, we explain this update-procedure line by line.
For each possible process state $X$, Algorithm \ref{alg:tensor} first checks whether the probability to be in $X$ is estimated to be greater than $0$ (lines 2-3).
If it is, $T^\mathrm{AoII}$ is updated for every state $X'$ at the receiver (line 4).

As the \gls{aoii} increases by $1$ whenever the information at the receiver is wrong, in lines 5-8, all distributions are shifted to the right and stored in a temporary variable $R$.

In lines 9-19, the \gls{aoii} distribution for every process state $X''$ and state $X'$ at the receiver is updated, whenever the estimated probability for the observed process to be in state $X''$ is positive.

If $X'' = X'$, the information at the receiver is correct, the \gls{aoii} is $0$ and the respective \gls{aoii} distribution is set accordingly (line 18-19).

If $X'' \neq X'$, $T_{t}^{\mathrm{AoII}}$ is increased by the probability of the observed process to evolve from $X$ in time step $t$ to $X''$ in time step $t$ weighted by $R(\Delta)$ for every \gls{aoii} $\Delta$ (lines 12-17).

Finally, the algorithm returns the updated tensor $T^\mathrm{AoII}_{t}$ in line 20.
\subsubsection{\rev{Time complexity of Algorithm \ref{alg:tensor}}}
\rev{In the worst case, Algorithm \ref{alg:tensor} iterates over all previous process states $X$, all receiver states $X'$, all current process states $X''$, and all age values $\Delta \in \{0,\ldots,M\}$. Therefore, its time complexity is in $\mathcal{O}(|\mathcal{X}|^3 M)$.
}

\subsection{Cost Prediction}
\label{sec:cost_prediction}

\subsubsection{\rev{Intuition behind Algorithm \ref{alg:cost}}}
\rev{Alg. 3 computes an expected one-step \valname degradation cost for the current belief. 
In other words, it marginalizes over all process states, states at the receiver, and age values that remain possible under the belief.}

\subsubsection{\rev{Description of Algorithm \ref{alg:cost}}}
If the age-based metric $F$ used as input for the \gls{got} metric is the \gls{aoi}, the expected cost is calculated as specified in line 2.
If $F$ is the \gls{aoii}, the expected cost is obtained as specified in line 4.

In both cases, every possible combination of states $X$ of the process, $X'$ at the receiver, and ages $\Delta$ is considered by summing over the belief distribution weighted by the goal-oriented tensor.
\begin{algorithm}[t]
\caption{\textit{PredictCost}}
\label{alg:cost}
\DontPrintSemicolon
\SetKwInOut{Input}{Input}
\SetKwInOut{Output}{Output}
\begin{fivept}
\Input{Belief $B_t$, Goal-oriented Tensor $\mathrm{GoT}$}
\Output{Predicted Cost}
\uIf{$F = \mathrm{AoI}$}{
\[
    \hspace{-0.3cm} \mathrm{cost} \gets \hspace{-0.3cm} \sum_{\substack{
    X \in \mathcal{X} \\
    X' \in \mathcal{X} \\
    \Delta=0,\ldots,M
    }} \hspace{-0.4cm} 
    D^{X}_t(X)  D^{X^\mathrm{Rx}}_t(X')  D^{\mathrm{AoI}^\mathrm{Rx}}_t(\Delta)  \mathrm{GoT}(X,X',\Delta)
\]
}
\ElseIf{$F = \mathrm{AoII}$}{
    \[
    \hspace{-0.3cm} \mathrm{cost} \gets \hspace{-0.3cm} \sum_{\substack{
    X \in \mathcal{X} \\
    X' \in \mathcal{X} \\
    \Delta=0,\ldots,M
    }} \hspace{-0.4cm} 
    D^{X}_t(X)  D^{X^\mathrm{Rx}}_t(X')  (T^{\mathrm{AoII}}_t \cdot \mathrm{GoT})(X,X',\Delta)
\]
}
\Return $\mathrm{cost}$\;
\end{fivept}
\end{algorithm}

\subsubsection{\rev{Time complexity of Algorithm \ref{alg:cost}}}
\rev{Algorithm \ref{alg:cost} evaluates the expected \gls{got} cost by summing over all combinations of process state, receiver state, and age value. Therefore, for both $F=\mathrm{AoI}$ and $F=\mathrm{AoII}$, its time complexity is in $\mathcal{O}(|\mathcal{X}|^2 M)$.}

\subsection{\sysname}
\label{sec:psbo}
\begin{algorithm}[t]
\caption{\sysname}
\label{alg:approach}
\DontPrintSemicolon
\SetAlgoLined
\SetKwInOut{Input}{Input}
\SetKwInOut{Output}{Output}
\begin{fivept}
\Input{
    Belief $B_t$,\\
    max\_sleep (max. sleep duration),\\
    max\_tx\_steps (max. no of Tx steps),\\
    $w_e$,$w_\mathrm{\valindex}$,$T^w, P^w$, $T^s$, $P^s$,$P^{a}$,$P^{d}$, $P^i$.
}
\Output{Optimal sleep duration}

$\mathrm{best} \gets 0$\;
$\mathrm{acc\_dat\_cost} \gets 0$\;
$\mathrm{prev\_avg\_cost} \gets \infty$\;
$N_t^\mathrm{sleep} \gets \infty$\;
$\overline{B} \gets B_t$\;

\While{$\mathrm{best} < \mathrm{max\_sleep}$}{
    $\mathrm{total\_time} \gets \mathrm{best}$\;
    $\mathrm{avg\_cost} \gets  \frac{\mathrm{acc\_dat\_cost}}{\mathrm{total\_time}}$ \;
    $\overline{B'} \gets \overline{B}$\;
    \For{$\mathrm{tx\_step} \in \{1, \dotsc,\mathrm{max\_tx\_steps} \}$}{
        $p \gets (1-r_t) \cdot r_t^{\mathrm{tx\_step} - 1}$\;
        $\overline{B'} \gets \mathit{BeliefUpdate}(\overline{B'})$\;
        $\mathrm{dat\_cost} \gets w_\mathrm{\valindex} \cdot \mathit{PredictCost}(\overline{B'})$\;
        $\mathrm{e\_cost} \gets w_\mathrm{e} \cdot (T^\mathrm{w}P^\mathrm{w} + T^\mathrm{s} P^\mathrm{s} + T_{t + \mathrm{best} + \mathrm{tx\_step}}^\mathrm{a} P^\mathrm{a} + T^\mathrm{i}P^\mathrm{i} + T_{t + \mathrm{best} + \mathrm{tx\_step}}^\mathrm{d}P^\mathrm{d})$\;
        $\mathrm{avg\_cost} \gets \frac{\mathrm{total\_time} \cdot \mathrm{avg\_cost} + p \cdot (\mathrm{dat\_cost} + \mathrm{e\_cost})}{\mathrm{total\_time} + p}$\;
        $\mathrm{total\_time} \gets \mathrm{total\_time} + p$\;
    }

    \uIf{$\mathrm{avg\_cost} > \mathrm{prev\_avg\_cost}$}{
        $\mathrm{best} \gets \mathrm{best} - 1$\;
        \textbf{break}\;
    }

    $\mathrm{prev\_avg\_cost} \gets \mathrm{avg\_cost}$\;
    $\mathrm{best} \gets \mathrm{best} + 1$\;
    $\overline{B} \gets \mathit{BeliefUpdate}(\overline{B})$\;
    $\mathrm{acc\_dat\_cost} \gets \mathrm{acc\_dat\_cost} + \mathit{PredictCost}(\overline{B})$\;
}

$N_t^\mathrm{sleep} \gets \mathrm{best}$\;

\Return $\mathrm{best}$\;
\end{fivept}
\end{algorithm}

\subsubsection{\rev{Intuition behind Algorithm \ref{alg:approach}}}
Algorithm \ref{alg:approach} determines the optimal deep-sleep duration through a simulation-based optimization of the expected time-averaged cost\footnote{A full python-based implementation can be found under https://github.com/wanjads/SatelliteIoTRL.}.

In ascending order, the algorithm considers possible deep-sleep durations and estimates the corresponding expected time-averaged cost based on the device’s current belief.
These estimates are high for short deep-sleep durations, because of high energy cost. 
As the deep-sleep duration increases, the total cost initially decreases because the energy cost drops. 
Beyond a certain point, however, longer deep-sleep periods start to degrade the \valname at the receiver. 

Once the cost of \valname degradation becomes larger than the decrease in energy cost, the overall cost begins to increase again.
At this point, the algorithm has found the optimal duration for the upcoming deep-sleep.
In the following, we explain the steps of this procedure line by line.

\subsubsection{\rev{Description of Algorithm \ref{alg:approach}}}
As inputs, \sysname needs all parameters concerning energy consumption and cost, namely $w_e$, $w_\mathrm{\valindex}$, $T^w, P^w$, $T^s$, $P^s$,$P^{a}$,$P^{d}$,and $P^i$, as well as a maximum deep-sleep duration, the current belief at the device, and a maximum number of steps used for transmission attempts after waking up.

Lines 1–5 initialize variables: the optimal deep-sleep duration $\mathrm{best}$, accumulated \valname degradation cost $\mathrm{acc\_dat\_cost}$, previously computed expected time-averaged cost $\mathrm{prev\_avg\_cost}$, the deep-sleep duration counter $N_t^\mathrm{sleep}$, and the current belief $\overline{B}$.
Here, we use a $\overline{B}$ for the device's belief during the simulation. 

Lines 6–25 iterate to find the optimal deep-sleep interval.
The loop is terminated if (a) this optimal deep-sleep interval is found (line 20) or (b) a maximal sleep duration $\mathrm{max\_sleep}$ is reached (line 6).
In the latter case, the \gls{iot} device sleeps for $\mathrm{max\_sleep}$ steps, i.e., $N_t^\mathrm{sleep} = \mathrm{max\_sleep}$.

Within each iteration, lines 7-9 are used to initialize the current total duration $\mathrm{total\_time}$, the expected time-averaged cost $\mathrm{avg\_cost}$, and to copy the current belief $\overline{B}$ to $\overline{B'}$.
Starting from this belief $\overline{B'}$, the simulation for the transmission steps after the deep-sleep interval of duration $\mathrm{best}$ starts in line 10.

We simulate a predefined number of $\mathrm{max\_tx\_steps}$ transmission time steps.
During each transmission step, we first calculate the probability that the device needs exactly $\mathrm{tx\_step}$ time steps for the transmission in line 11.

At this point, the algorithm uses $r_t = C^\mathrm{data}_\mathrm{erasure,t} \cdot C^\mathrm{feedback}_\mathrm{erasure,t}$, which is the probability to receive an ACK after a transmission in time step $t$.
To estimate $r_t$, the \gls{iot} device tracks previous transmission attempts and calculates a mean by dividing the number of successful transmission attempts by the total number of transmission attempts.
Given $r_t$, the probability that the transmission succeeds after exactly $\mathrm{tx\_step}$ time steps is $(1-r_t) \cdot r_t^{\mathrm{tx\_step} - 1}$.

Line 12 updates the belief $\overline{B'}$ used during the simulation of the transmission phase.
In line 13, the cost $c_\mathrm{\valindex}$ resulting from \valname degradation in the current transmission time step are predicted through Algorithm \ref{alg:cost}.
Line 14 computes the energy cost ($\mathrm{e\_cost}$) of waking up, sensing, and transmitting, weighted by $w_\mathrm{e}$ according to Eq. \ref{eq:energy_costs}.

The probability $P^\mathrm{Tx}(\overline{B})$ to transmit given belief $\overline{B}$ depends on the metric $F$.
The average required time $T_t^\mathrm{a}$ during which the antenna is powered during transmission time step $t$ is approximated by
\begin{equation}
    T_t^\mathrm{a} = \min(T, r_t \cdot \overline{C_\mathrm{delay}} + \frac{(1-r_t)\cdot T^\mathrm{a}}{r_t}),
\end{equation}
where $\overline{C_\mathrm{delay}} = \overline{C^\mathrm{data}_\mathrm{delay}} + \overline{C^\mathrm{feedback}_\mathrm{delay}}$ is the average delay between the transmission of a measurement and the reception of an ACK at the device, which the \gls{iot} device estimates as a mean from round-trip times obtained through feedback from previous transmission attempts.

In line 15, the expected time-averaged cost \textit{avg\_cost} is computed.
Line 16 increases the total time by an additional expected duration $p$, which is the number of transmission time steps that are required on average to successfully transmit.

Lines 18–20 implement a convexity check: once the expected time-averaged cost begins to increase, the previous deep-sleep duration is optimal, assuming the entries in the \gls{got} are monotone in the sense that as long as the device does not transmit after some time step $t_0 \in \mathbb{N}$
\begin{align*}
    &\forall t, t' > t_0, X,X',X'',X''' \in \mathcal{X}: \\
    &t < t' \Rightarrow \mathrm{GoT}(X,X',F_t) \leq \mathrm{GoT}(X'',X''',F_{t'}).
\end{align*}

A weaker and still sufficient assumption for optimality is that the sequence of expected data degradation costs $c_\mathrm{\valindex}$ is convex.
If none of these conditions is satisfied, longer deep-sleep durations with lower expected time-averaged costs may exist. 
In this case, our approach conservatively returns the first local minimum.

After finding the minimum, the algorithm terminates.
Otherwise, the loop continues, updating $\mathrm{prev\_avg\_cost}$, $\mathrm{best}$, $\overline{B}$, and $\mathrm{acc\_dat\_cost}$ accordingly (lines 21–24).

Finally, the algorithm sets the optimal sleep duration $N_t^\mathrm{sleep}$ in line 26 and returns it in line 27.

\subsubsection{\rev{Time complexity of Algorithm \ref{alg:approach}}}
\rev{In the worst case, Algorithm \ref{alg:approach} performs $\mathrm{max\_sleep}$ outer iterations. 
In each outer iteration, it simulates up to $\mathrm{max\_tx\_steps}$ transmission steps, and each simulated step calls Algorithm \ref{alg:update} and Algorithm \ref{alg:cost} once.
Hence, the worst-case time complexity of \sysname is in $\mathcal{O}(\mathrm{max\_sleep} \cdot  \mathrm{max\_tx\_steps} \cdot |\mathcal{X}|^2 \cdot M)$ for $F=\mathrm{AoI}$ and in $\mathcal{O}(\mathrm{max\_sleep} \cdot  \mathrm{max\_tx\_steps} \cdot |\mathcal{X}|^3 \cdot M)$ for $F=\mathrm{AoII}$. 
Thus, \sysname scales linearly in the search horizon and polynomially in the size of the process state space and the age cap. 
The dominant term in the \gls{aoii} case is the tensor update in Algorithm \ref{alg:tensor}.}

\section{HARDWARE IMPLEMENTATION}
\label{sec:implementation}

To validate the applicability of the proposed \sysname approach under realistic conditions, a status update system is implemented.

The physical setup follows the structure of the system model from Section \ref{sec:system_model}, with one additional component, namely, the control station (a Linux-based laptop running a Python script) which replaces the physical sensing process to allow reproducible experiments. 
The control station additionally serves as a logging interface. 

\begin{figure}[!t]
    \centering
    \includegraphics[width=0.95\linewidth]{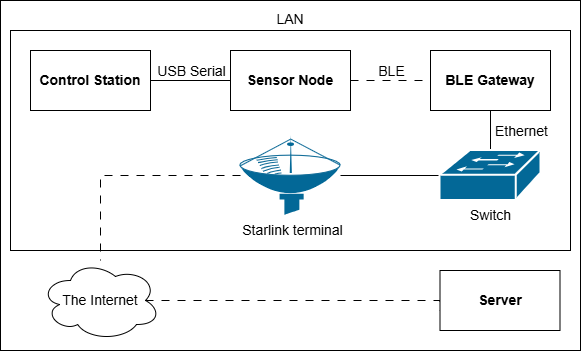}
    \caption{\rev{Scheme of the physical layout of the experimental setup.}}
    \label{fig:setup_scheme}
    \vspace{-0.3cm}
\end{figure}

\begin{figure}[!t]
    \centering
    \includegraphics[width=0.95\linewidth]{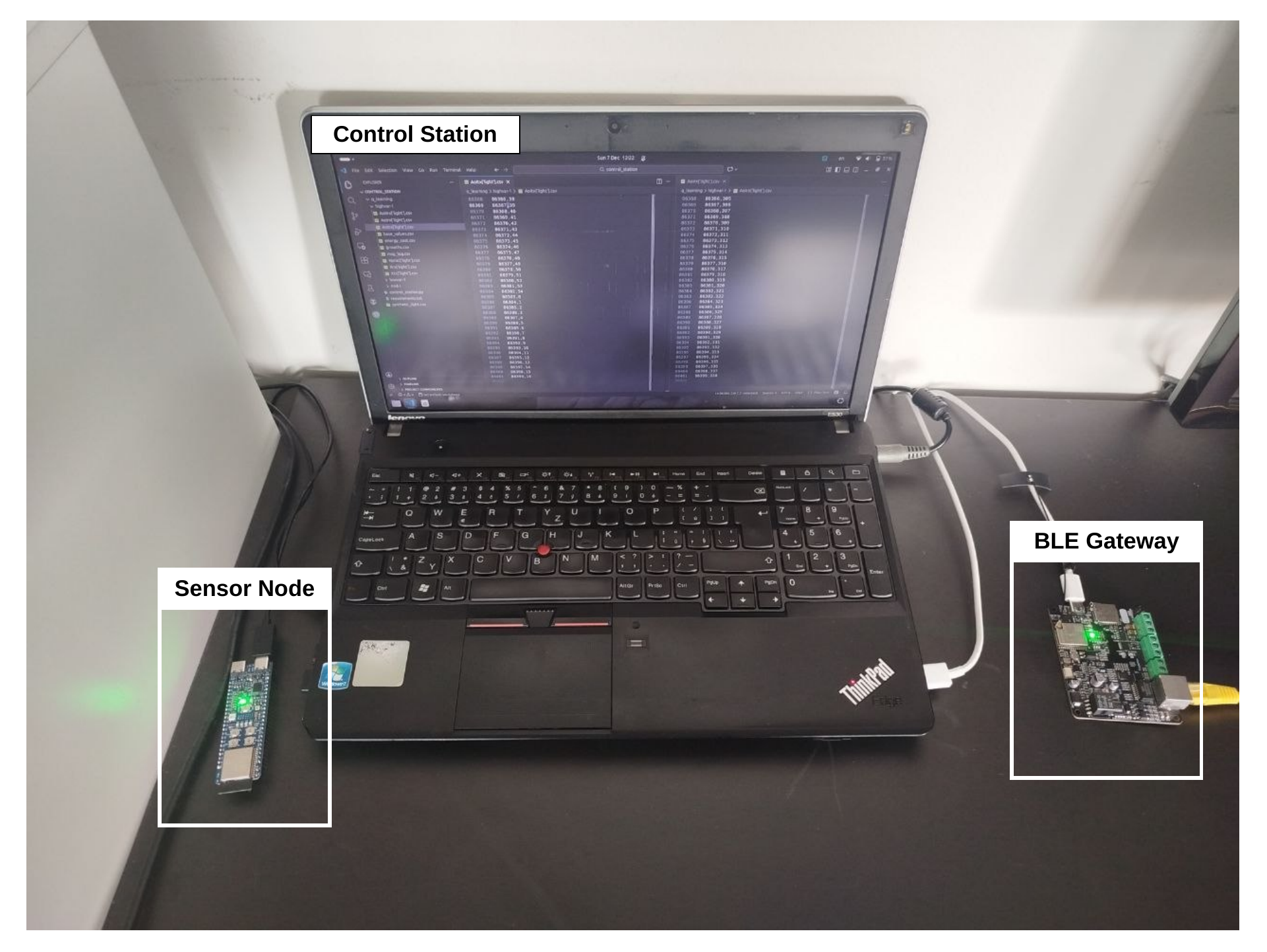}
    \caption{Physical layout of the experimental setup consisting of the control station, sensor node, and BLE gateway.}
    \label{fig:setup_photo}
    \vspace{-0.3cm}
\end{figure}

Apart from this substitution, the interactions between elements reproduce the logical flow of the modeled status update system.
Apart from the control station, we use a sensor node transmitting to a Bluetooth Low Energy (BLE) gateway, which receives measurements from the sensor node and is connected to a Starlink user terminal. 
This user terminal is used to connect the setup via the Starlink constellation to a remote server (a Linux host running a Python script). 
\rev{Fig. \ref{fig:setup_scheme} shows the physical layout of the experimental setup. 
The \gls{iot} device was deployed in Bratislava, Slovakia, while the remote server was hosted in Germany.
A photo of the sensor node, the control station and the BLE gateway is shown in Fig. \ref{fig:setup_photo}.
The BLE gateway is connected to the Starlink user terminal, which is not shown in the photo.}

\rev{Compared to a fully deployed sensing application, the experimental setup mainly differs in that the sensing process is replaced by a control station replaying prerecorded traces.
Apart from this substitution, the communication which is modeled as a single Markov channel in the theoretical model is now implemented using a real \gls{siot} setup including the BLE gateway, a Starlink user terminal and the Starlink network.
PSBO is implemented on the sensor node itself, while the control station and remote server only support trace replay, logging, and communication.}

The hardware implementation of the experimental system relies on widely used ESP32-based development boards and standard computing devices. 
The sensor node is built on the FIIT ESP32 Kit \cite{Poncak}, a custom development board designed at the Slovak University of Technology in Bratislava (FIIT STU) for research in \gls{iot} applications.

In the theoretical model, the \gls{iot} device measures the environmental process directly. 
In the experimental setup, the corresponding measurements were taken offline, are stored on the control station and transmitted to the sensor node over a USB serial link. 
The node treats the received value as its current observation $X_t$, stores it as $X_t^{Tx}$, and uses it for belief updates and decision-making as described in previous sections.

\rev{The used measurements are light-intensity values collected during one hour of car driving around Bratislava. 
We use light intensity here as one concrete example of the generic environmental process introduced in Section III. 
The purpose of this trace is to provide a real time-varying sensed signal for the hardware proof of concept.}

\rev{The one-hour trace was chosen as a practical measurement record that provides realistic temporal variation. 
It is used to estimate the parameters of the Markov chain, from which a 24-hour time horizon is generated.}
Using externally provided traces makes it possible to evaluate different agents under identical conditions, since all of them operate on the same sequence of environmental values. 

The \gls{iot} device used in the implementation is an ESP32-S3-based microcontroller, which provides the required interfaces for serial communication, BLE transmission, and deep-sleep operation. 
These capabilities allow the device to implement the behavior defined in Section \ref{sec:system_model}.

The communication chain corresponds to the multi-hop link in Fig. \ref{fig:systemmodel}, with additional communication between the control station and the sensor node. 

The BLE gateway is implemented on an ESP32-based development board that provides both BLE and Ethernet connectivity. 
BLE enables the short-range connection between the sensor node and the gateway, while the gateway forwards packets to the remote server via an Ethernet link routed through the Starlink network. 
Acknowledgment messages allow the sensor node to update its beliefs about $X^{\mathrm{Rx}}$, $\mathrm{AoI}^{\mathrm{Rx}}$, $\mathrm{AoII}$ and the channel behavior based on observed outcomes.

The firmware running on the microcontrollers is implemented in C/C++ using the \textit{PlatformIO} development environment. 
It uses the \textit{Arduino framework} built on top of the \textit{ESP-IDF} environment, combining high-level Arduino APIs with low-level ESP32 functionality. 
The Arduino framework is primarily used for serial communication and timing functions, while the ESP-IDF headers provide access to BLE stack configuration and control. 

The firmware is structured as a main event loop with asynchronous callbacks for BLE and timer events, executing on the FreeRTOS-based ESP-IDF runtime. 

\rev{Table \ref{tab:hw_sw_specs} summarizes the hardware and software 
specifications of the IoT device and of the BLE gateway used in the experimental setup.}

\begin{table}[!t]
\color{tudablue}
\scriptsize
\centering
\renewcommand{\arraystretch}{1.2}
\begin{tabular}{lll}
\toprule
\textbf{Parameter} & \textbf{Sensor Node} & \textbf{BLE Gateway} \\
\midrule
\multicolumn{3}{l}{\textit{Hardware}} \\
Board           & FIIT ESP32 Kit (FIIT STU) & Custom ESP32 board \\
Module          & ESP32-S3-WROOM-1-N16R2 & ESP32-WROOM-32UE-N16 \\
SoC             & ESP32-S3 & ESP32 \\
CPU             & Xtensa LX7, dual-core  & Xtensa LX6, dual-core \\
\makecell[l]{Clock\\frequency} & 240\,MHz & 240\,MHz \\
Flash memory    & 16\,MB & 16\,MB \\
PSRAM           & 2\,MB & --- \\
Bluetooth       & BLE 5.0 & BLE 4.2 \\
Antenna         & PCB (integrated) & External (U.FL) \\
\midrule
\multicolumn{3}{l}{\textit{Software}} \\
Framework & \makecell[l]{ESP-IDF \\ (Arduino for serial/timing)}  & \makecell[l]{ESP-IDF \\ (Arduino for serial/timing)} \\
\makecell[l]{ESP-IDF\\version} & v4.4.7 & v4.4.7 \\
\makecell[l]{FreeRTOS\\version} & v10.4.3 & v10.4.3 \\
Compiler        & \makecell[l]{GCC 12.2.0 \\ (xtensa-esp32s3-elf)} & \makecell[l]{GCC 12.2.0 \\ (xtensa-esp32-elf)} \\
IDE             & PlatformIO & PlatformIO \\
Language        & C/C++ & C/C++ \\
\bottomrule
\end{tabular}
\caption{\rev{Hardware and Software Specifications}}
\label{tab:hw_sw_specs}
\end{table}
\section{EVALUATION}
\label{sec:simulations}

\begin{table}
\scriptsize
\centering
\renewcommand{\arraystretch}{1.9}
\begin{tabular}{c p{2.3cm} c c}
\hline
\rule{0pt}{2.6ex}
Parameter & \makecell{Description} & Value & Range \\
\hline
\hline
\rule{0pt}{2.6ex}
$N$ & \makecell{number of\\ repetitions} & $100$ & - \\
\hline
$t_\mathrm{final}$ & \makecell{number of \\ time steps \\ per repetition} & $86,400$ & - \\
\hline
$T$ & \makecell{time step duration} & $1$s & - \\
\hline
$T^\mathrm{Tx}$ & \makecell{transmission duration} & $64$ ms & - \\
\hline
$T^\mathrm{w}$ & \makecell{wake up duration} & $47$ ms & - \\
\hline
$T^\mathrm{s}$ & \makecell{sensing duration} & \makecell{$52$ ms (temperatures) \\ $0$ ms (light intensities)} & - \\
\hline
$C_\mathrm{data}^\mathrm{delay}$ & \makecell{data channel delay} & $40$ - $220$ ms & - \\
\hline
$C_\mathrm{feedback}^\mathrm{delay}$ & \makecell{feedback\\ channel delay} & $40$ - $220$ ms & - \\
\hline
$C_\mathrm{feedback}^\mathrm{erasure}$ & \makecell{feedback channel\\erasure rate} & $0.00$ & - \\
\hline
$w_\mathrm{\valindex}$ & \makecell{\valname weight} & $1$ & - \\
\hline
$|\mathcal{X}^d|$ & \makecell{number of states\\per process} & \makecell{$8$ (temperatures)\\ $7$ (light intensities)} & - \\
\hline
\rowcolor[gray]{0.9}
$p_\mathcal{X}$ & \makecell{state change rate} & $0.01$ & $2^{-9}$ to $2^{-2}$ \\
\hline
\rowcolor[gray]{0.9}
$C_\mathrm{data}^\mathrm{erasure}$ & \makecell{data channel \\ erasure rate} & $0.01$ & $0.00$ to $0.99$ \\
\hline
\rowcolor[gray]{0.9}
$w_\mathrm{e}$ & \makecell{energy weight} & $1 \frac{1}{\mathrm{J}}$ & $2^{-3}$ to $2^{-5}$ \\
\hline
\end{tabular}
\caption{\tuda{Simulation Parameters}}
\label{table:SimPara}
\vspace{-0.5cm}
\end{table}

\begin{figure*}[htb]
    \centering
    \begin{subfigure}[b]{0.48\textwidth}
        \centering
        \includegraphics[width=\columnwidth]{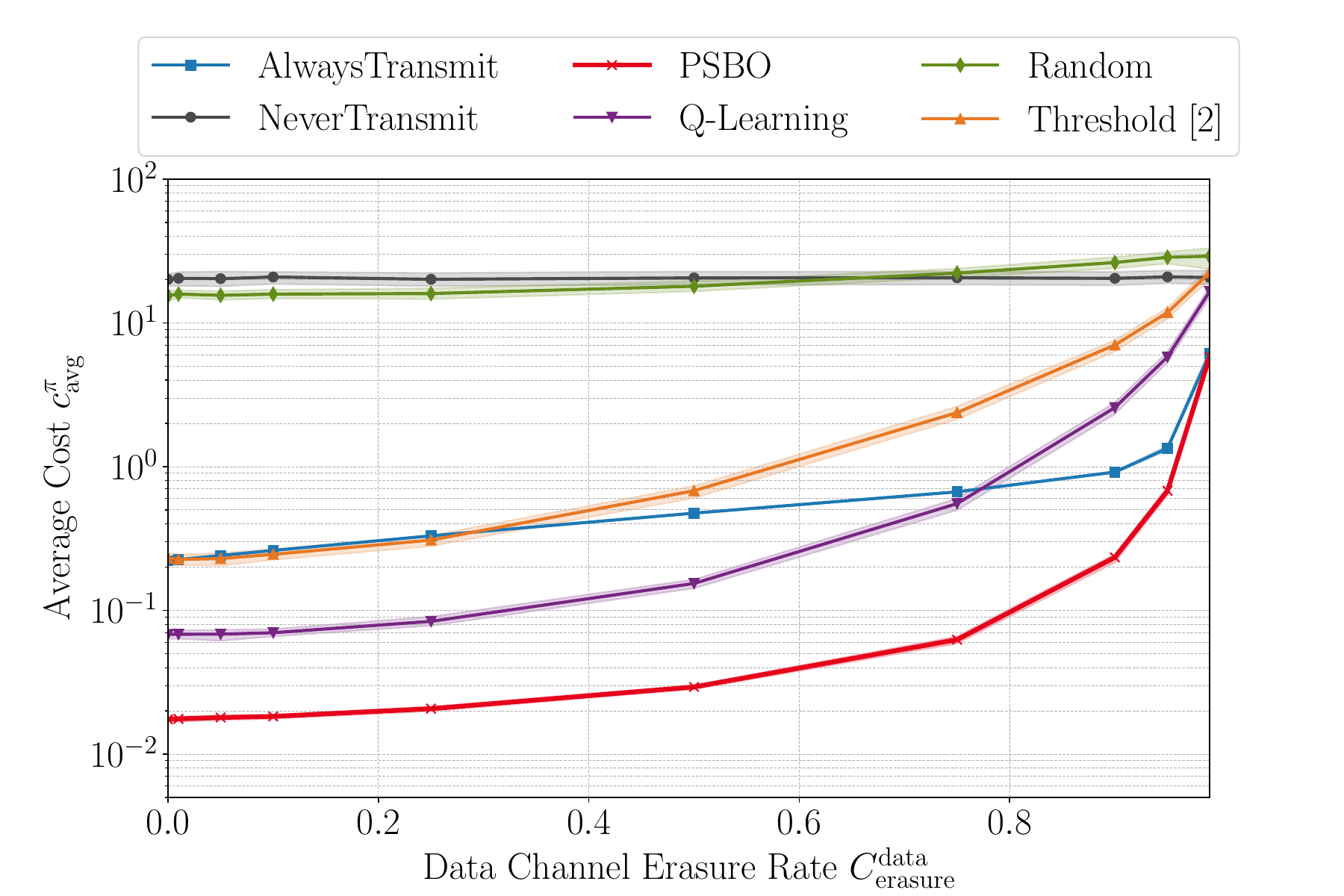}
        \caption{Varying data channel erasure rates}
        \label{fig:specific_drop_rate}
    \end{subfigure}
    \begin{subfigure}[b]{0.48\textwidth}
        \centering
        \includegraphics[width=\columnwidth]{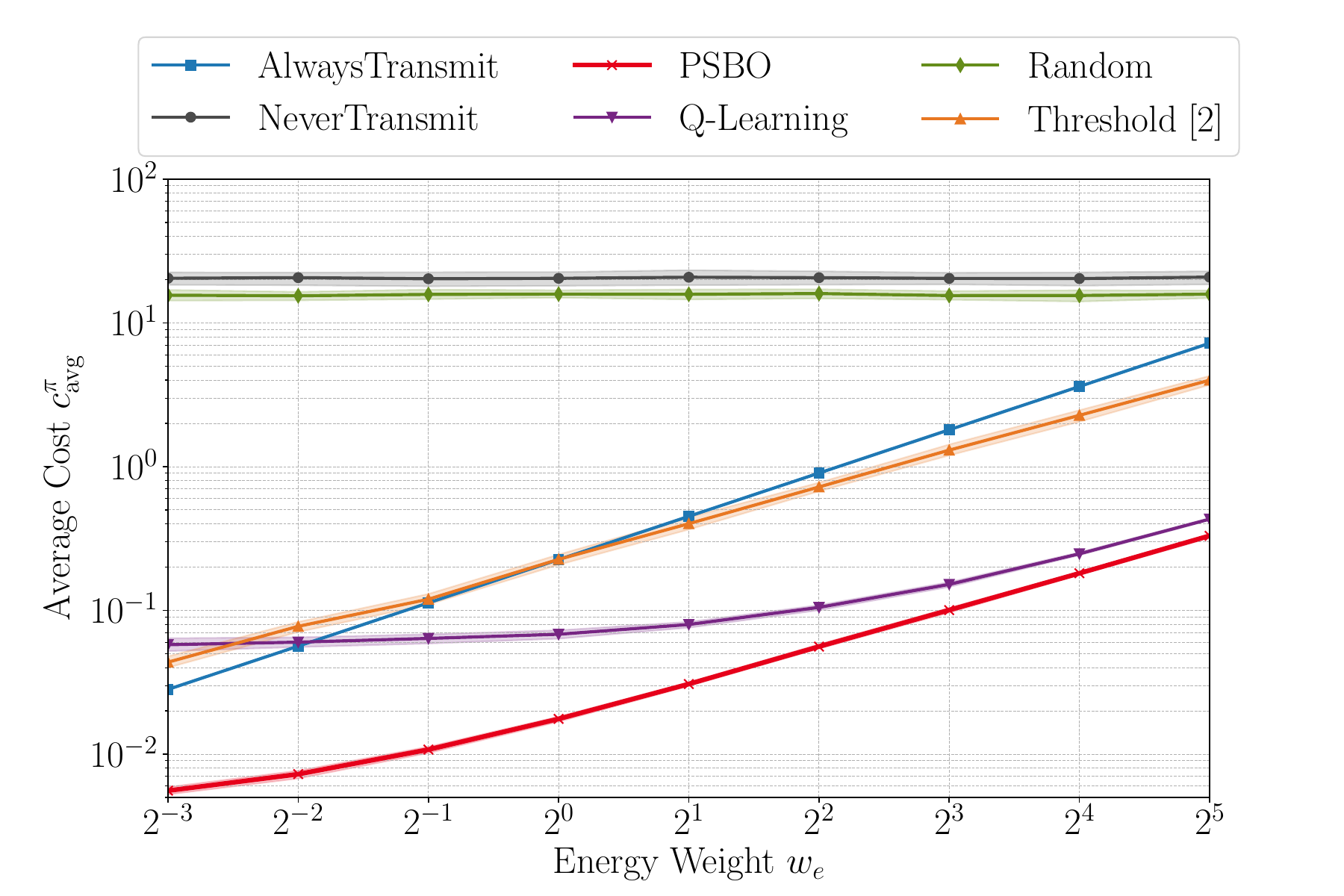}
        \caption{Varying energy weights}
        \label{fig:specific_energy_weight}
    \end{subfigure}
    \caption{Average cost of \sysname and baselines for varying parameters (\gls{got} \textbf{A})}
    \label{fig:specific}
\end{figure*}

\begin{figure}[htb]
    \centering
    \includegraphics[width=\columnwidth]{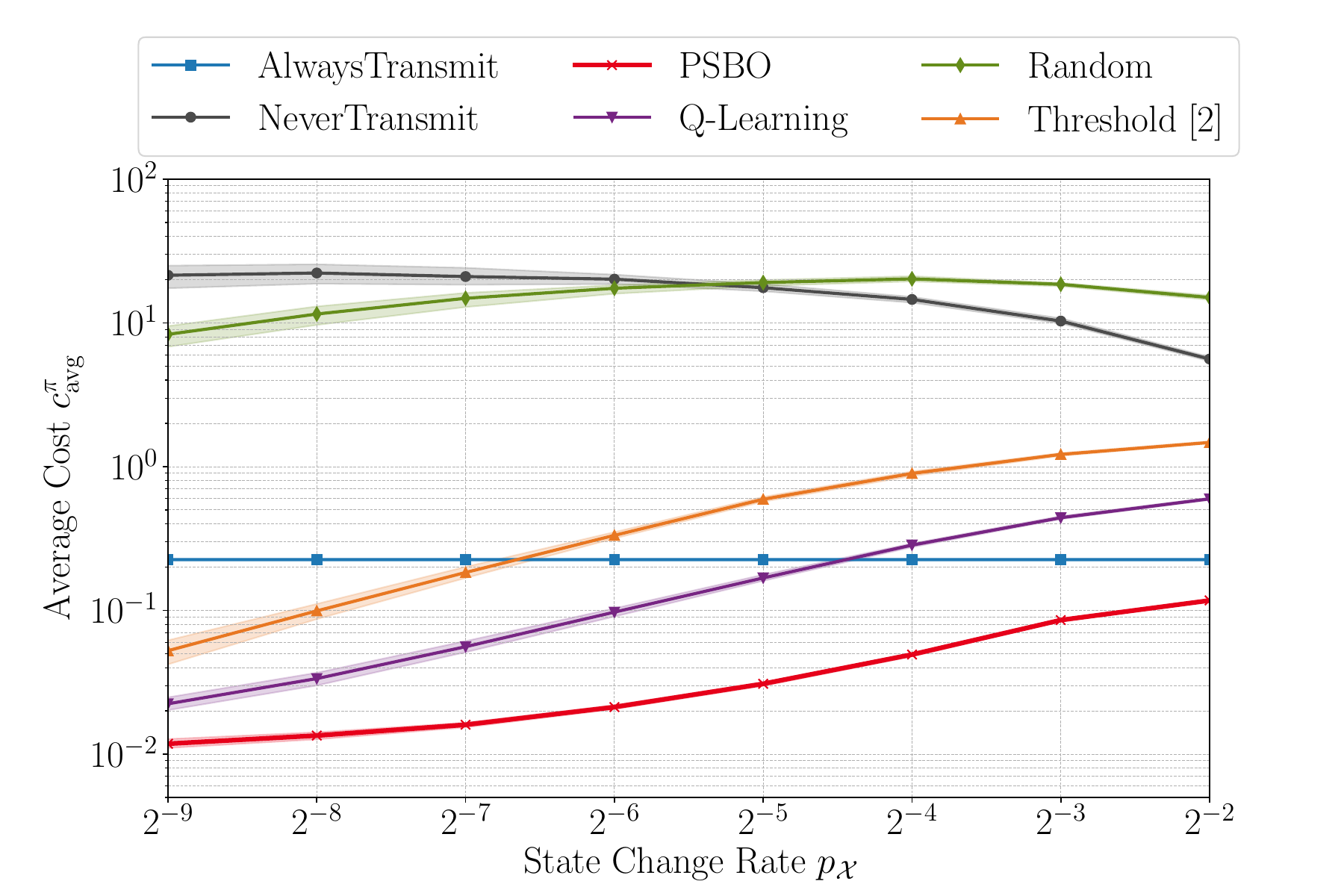}
    \caption{Average cost of \sysname and baselines for varying state change rates (\gls{got} \textbf{A})}
    \vspace{-0.5cm}
    \label{fig:specific_prob}
\end{figure}

\begin{figure*}[htb]
    \centering
    \begin{subfigure}[b]{0.48\textwidth}
        \centering
        \includegraphics[width=\columnwidth]{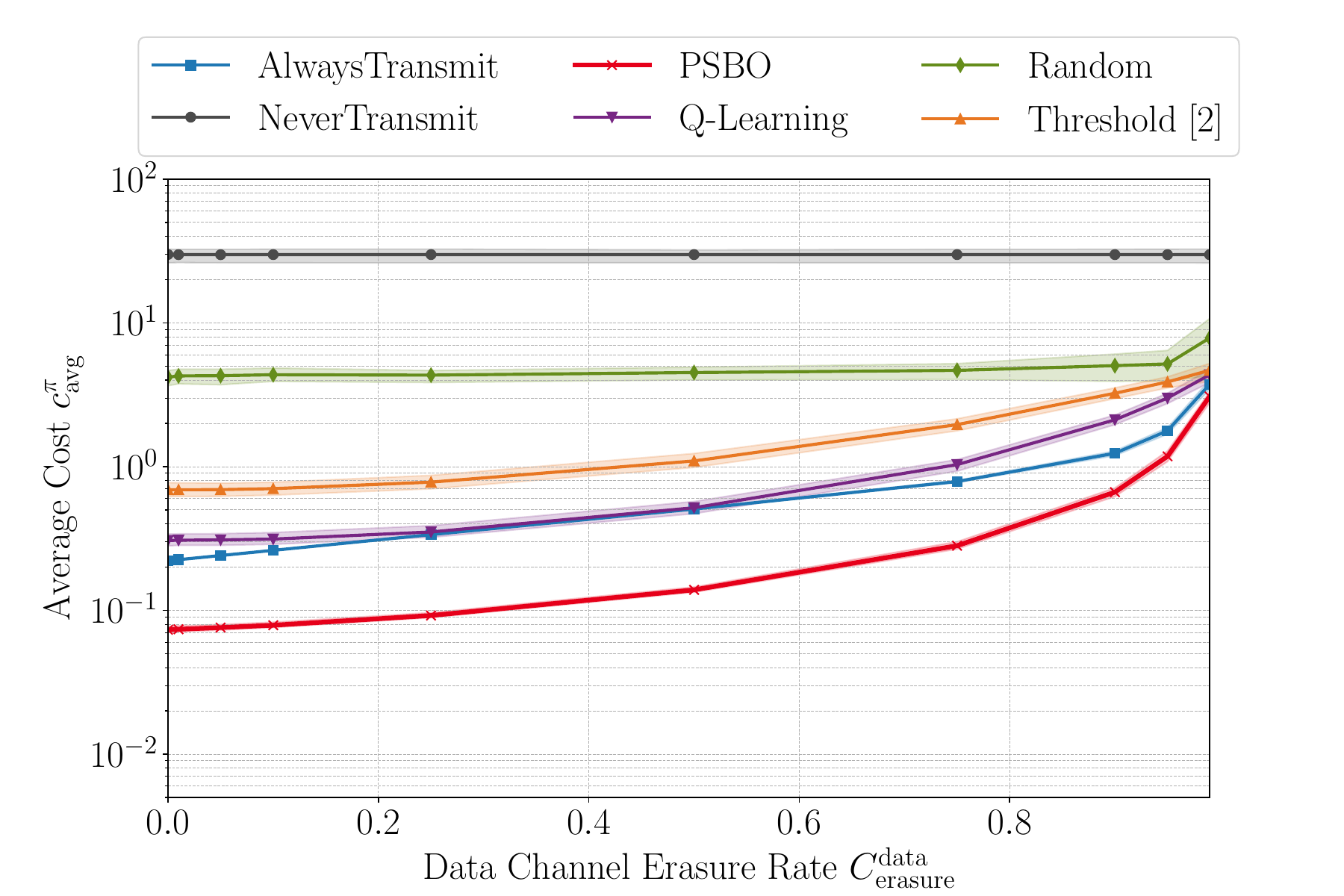}
        \caption{Varying data channel erasure rates}
        \label{fig:general_drop_rate}
    \end{subfigure}
    \begin{subfigure}[b]{0.48\textwidth}
        \centering
        \includegraphics[width=\columnwidth]{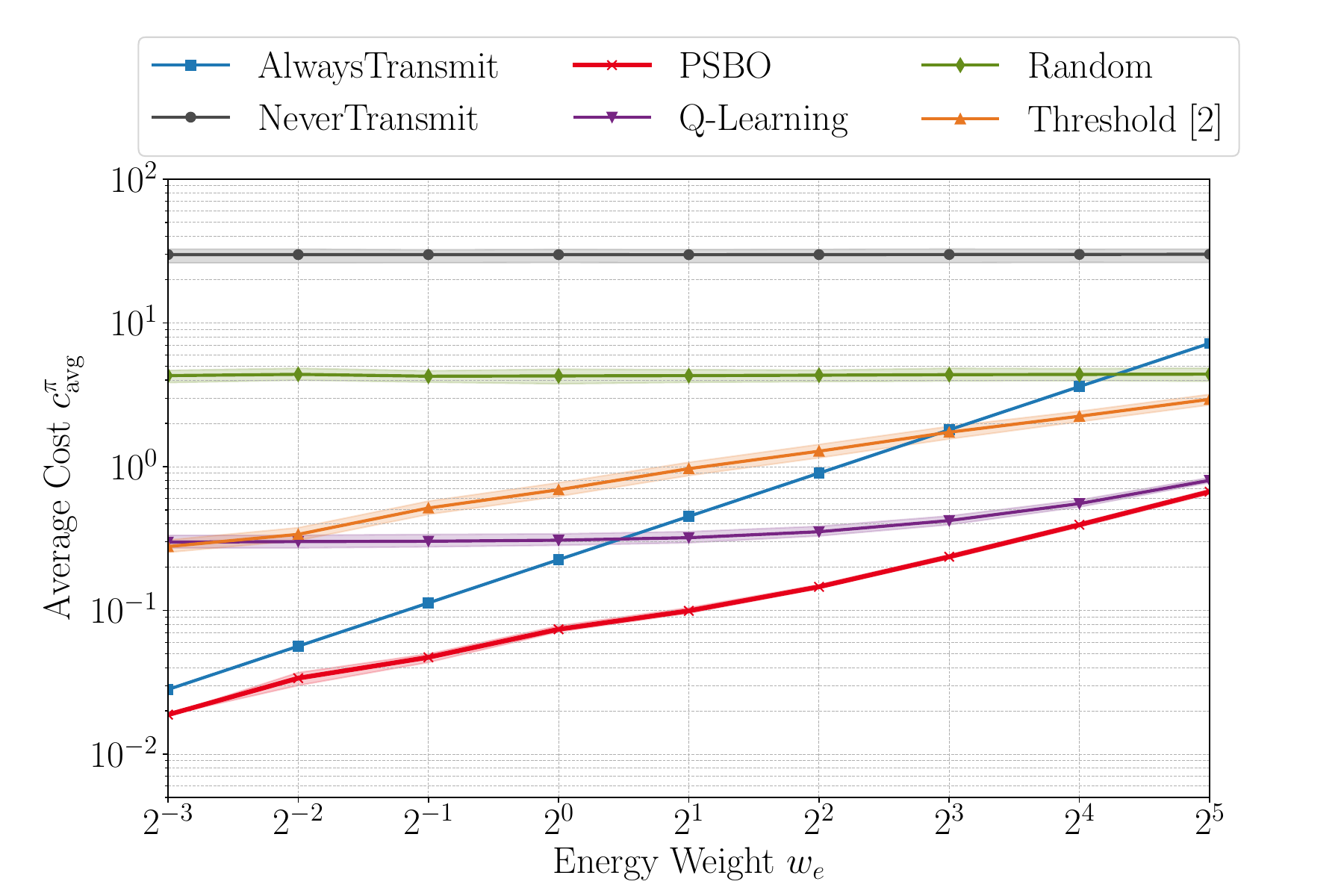}
        \caption{Varying energy weights}
        \label{fig:general_energy_weight}
    \end{subfigure}
    \caption{Average cost of \sysname and baselines for varying parameters (\gls{got} \textbf{B})}
    \label{fig:general}
    \vspace{-0.3cm}
\end{figure*}

\begin{figure*}[htb]
    \centering
    \begin{subfigure}[b]{0.32\textwidth}
        \centering
        \includegraphics[width=\columnwidth]{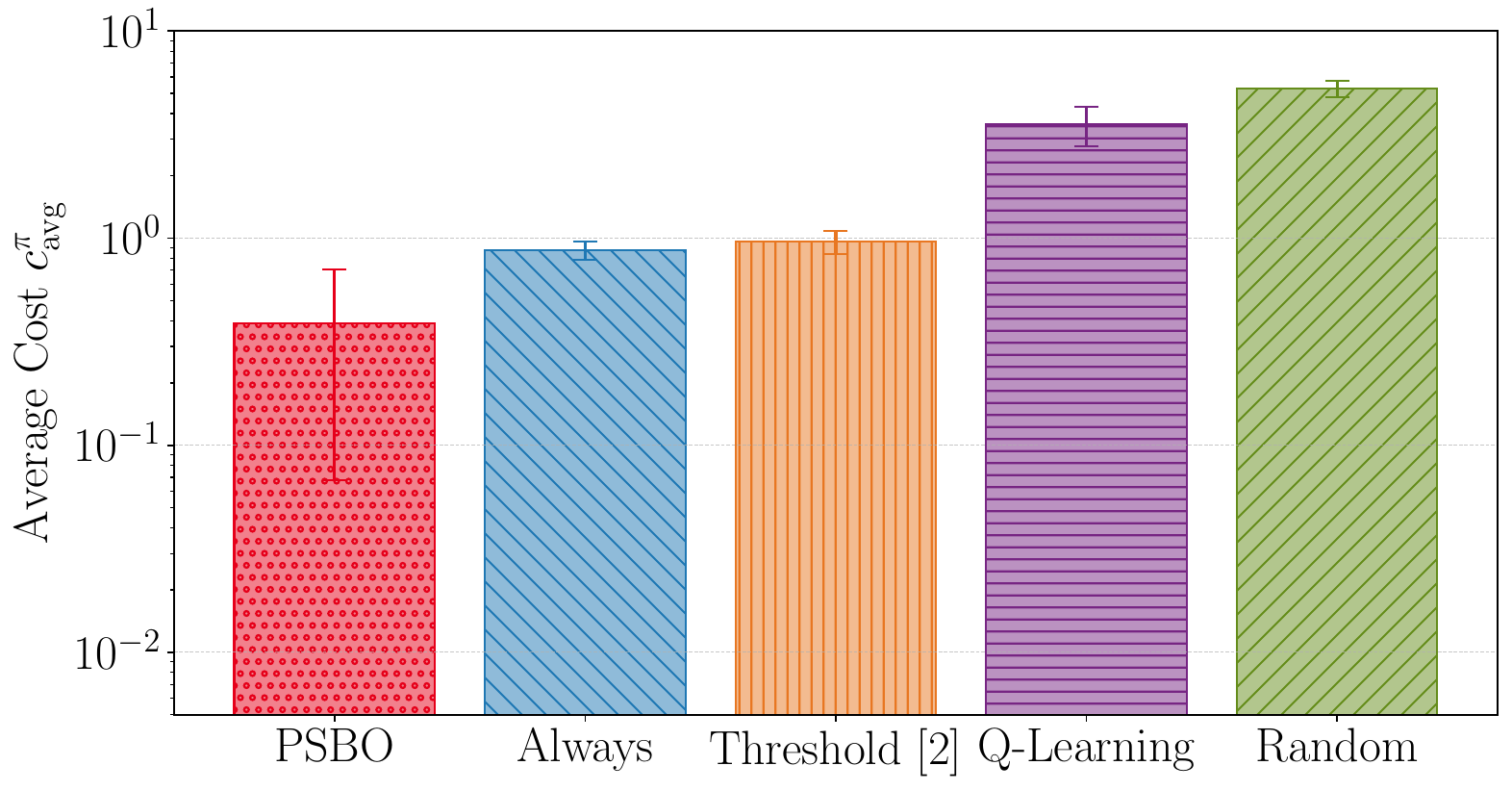}
        \caption{Comparison of average costs}
        \label{fig:experiments_costs}
    \end{subfigure}
    \begin{subfigure}[b]{0.32\textwidth}
        \centering
        \includegraphics[width=\columnwidth]{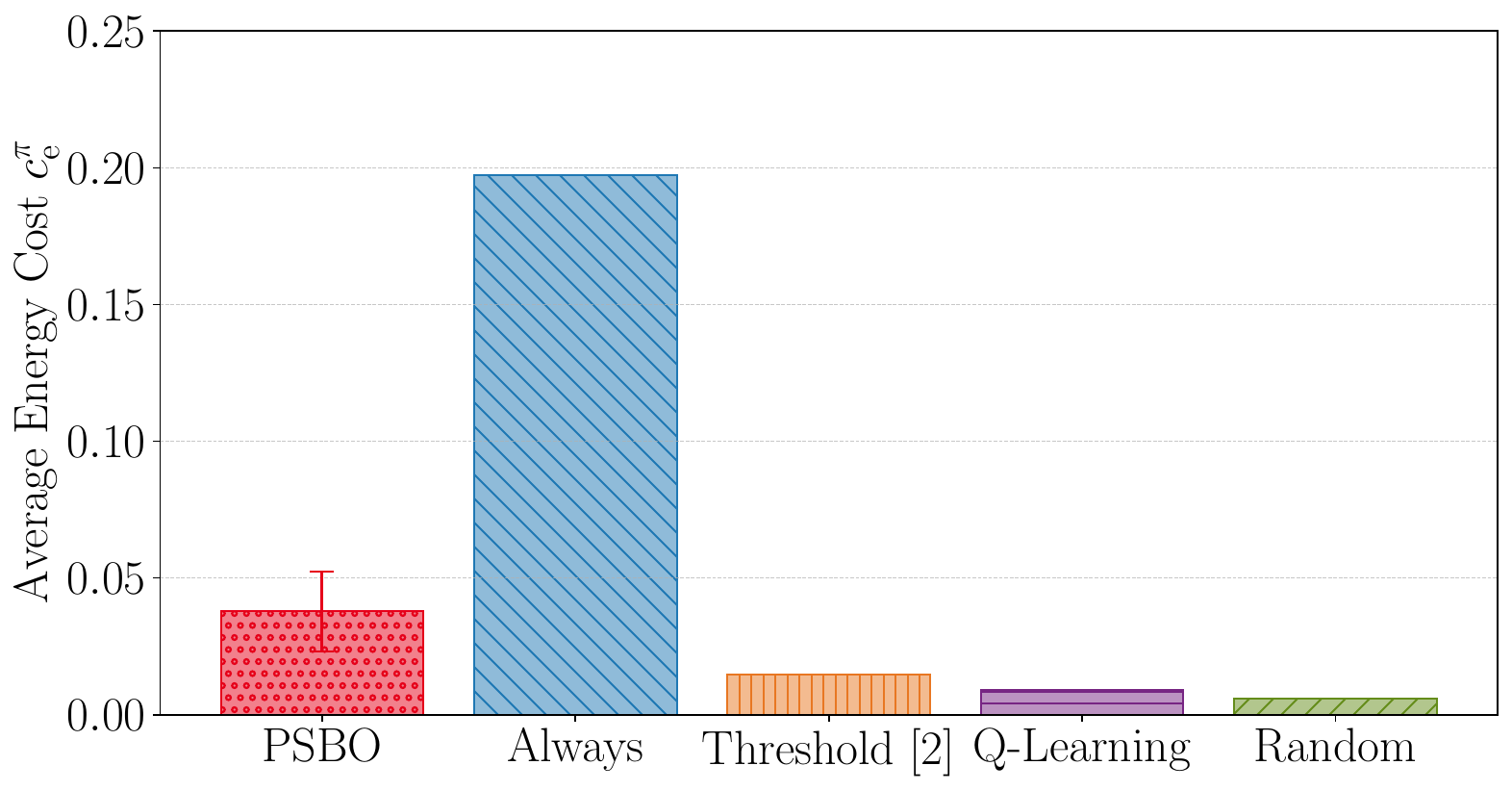}
        \caption{Comparison of average energy costs}
        \label{fig:experiments_energy}
    \end{subfigure}
    \begin{subfigure}[b]{0.32\textwidth}
        \centering
        \includegraphics[width=\columnwidth]{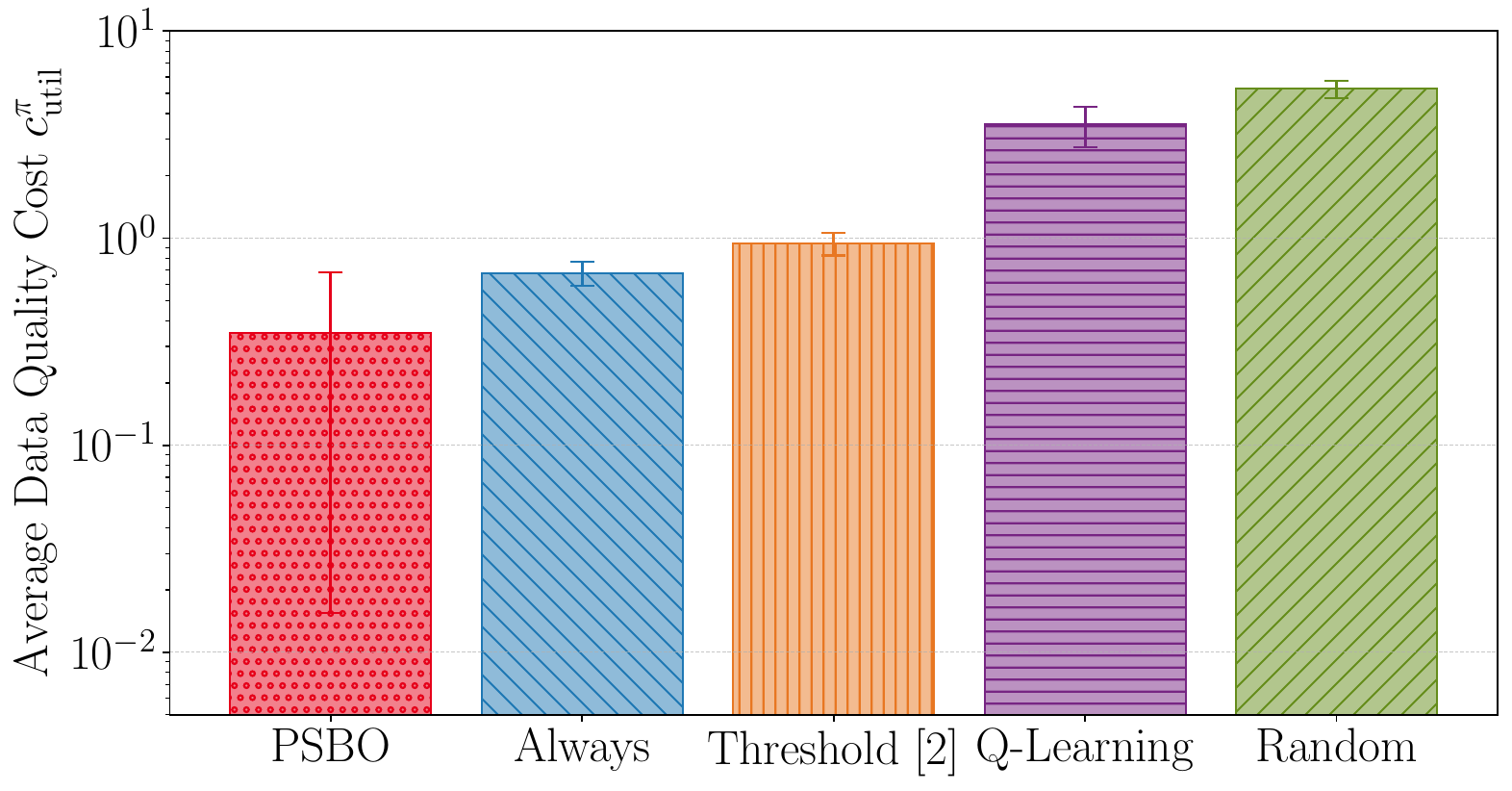}
        \caption{Comparison of average \valname costs}
        \label{fig:experiments_got}
    \end{subfigure}
    \caption{Experimental Results for our \sysname approach and baseline strategies}
    \label{fig:experiments}
    \vspace{-0.5cm}
\end{figure*}

In this section, we present a numerical and experimental evaluation of our \sysname approach against baseline strategies. Sec. \ref{sec:simulation_setup} describes the simulation parameters, while Sec. \ref{sec:energy_measurements} details the energy measurements used to set these parameters based on real hardware. Sec. \ref{sec:simulation_results} presents the simulation results, and Sec. \ref{sec:experimental_results} reports results of the experimental analysis.

\subsection{Simulation Setup}
\label{sec:simulation_setup}
In a first step, we simulated the model described in Sec. \ref{sec:system_model} numerically. Simulation parameters are listed in Table \ref{table:SimPara}.
The transition probabilities of the Markov chain $\mathcal{C}$ modeling the channel erasure rates and delays are derived from data obtained through the hardware setup described above in Sec. \ref{sec:implementation}. 
To this end, we logged a full day of delay values for every time step and quantized the resulting delays into $20$ms bins.
The occurrences of transitions between these bins then directly translate to the transition probabilities in $\mathcal{C}$.

We aim to cover a large span of possible parameters to analyze the performance of \sysname under diverse circumstances.

To this end, we vary the data channel erasure rate $C_\mathrm{data}^\mathrm{erasure}$ from $0.00$ to $0.99$. In the latter case, only $1$\% of transmission attempts succeed. 

The energy weight $w_\mathrm{e}$ takes values from $2^{-4}$, where \valname degradation is contributing a significantly larger share to the objective up to $2^{4}$, where energy consumption dominates the cost.

The process state change rate $p_\mathcal{X}$ describes how fast the observed process changes its state. 
We vary it from $2^{-9}$, which corresponds to processes that are nearly static and change every $8.5$ minutes on average to $2^{-2}$, which corresponds to processes that are highly volatile and change every $4$ seconds on average.

The varying parameters are listed in gray rows in Table \ref{table:SimPara}, where the values indicated correspond to the fixed values used when varying the other parameters.

To quantify \valname degradation, we use the \gls{got} metric.
To recall the meaning of \valname in this context, assume the simulation setting $w_{\mathrm{\valindex}} = 1,w_e = 1 \frac{1}{\mathrm{J}}$. Then, a \valname cost value of $1$ contributes one cost unit to the objective function and is therefore treated as equivalent to an energy cost of $1\,\mathrm{J}$.

We use two variants of the \gls{got}:
The first considered \gls{got}, \gls{got} \textbf{A}, partitions the process states into critical and non-critical ones. 
If the observed process is in a critical state $X_t$ while the receiver indicates a non-critical state $X_t^\mathrm{Rx}$, the resulting \valname is substantially lower than for any other inconsistency, and the corresponding costs increase linearly with \gls{aoii}.
All other state inconsistencies yield a \valname that decreases linearly with \gls{aoii} with a small scaling factor, here set to $0.001$.

With \gls{got} \textbf{B}, we illustrate that our approach can accommodate arbitrary definitions of \valname and is not dependent on patterns like in \gls{got} \textbf{A}.
To construct \gls{got} \textbf{B}, each state pair $(X_t, X_t^\mathrm{Rx})$ is assigned a random base \valname drawn uniformly from $[-0.5 - v,\,-0.5 + v]$, which is then decreased linearly with increasing \gls{aoii}.
The corresponding slope is scaled by an independent random factor uniformly distributed in $[0.5 - v,\,0.5 + v]$, where $v$ controls the variability of assigned costs. 

We evaluate \sysname against five baseline strategies:

The \textit{Random} strategy always decides to enter deep-sleep mode after the \gls{iot} device was awake for one time step. 
It chooses a deep-sleep duration uniformly at random between $30$ and $300$ seconds, independent of the system’s state. 

The \textit{AlwaysTransmit} strategy remains continuously active, sensing and transmitting at every time step.
This guarantees the highest \valname and results in the highest energy consumption.

Conversely, the \textit{NeverTransmit} strategy stays in deep-sleep mode for the entire time and completely omits communication.
This strategy minimizes energy consumption but suffers from minimal \valname.

The \textit{Threshold} strategy transmits only when a predefined \gls{aoi} threshold $\theta$ is exceeded. We choose $\theta$ such that this strategy optimizes the \gls{aoi}-case as derived in \cite{Cao2023}:
\begin{equation}
    \theta = \left\lfloor \sqrt{\frac{2 w_\mathrm{e} \cdot \left( T^s \cdot P^s + T^w \cdot P^w + T^{\mathrm{Tx}} \cdot P^a \right)}{w_{\text{\valindex}}}} \right\rfloor + 1
\end{equation}
This strategy is optimal under traditional \gls{aoi}-based metrics, but it neglects application-specific \valname and makes no use of context-specific information.

The \rev{\textit{Q-Learning}\footnote{\rev{Q-Learning is a classical model-free reinforcement learning method that learns a state-action-value table for discrete state-action pairs and uses it to select actions that minimize the expected long-term cost.}}} strategy adaptively selects sleep durations using reinforcement learning. 
It learns to minimize a long-term cost function that includes both energy and \valname.
To make tabular learning possible, we use discrete states for \textit{Q-Learning} which consist of $X^\mathrm{Tx}, X^\mathrm{Rx}, \mathrm{AoI}^\mathrm{Tx}, \mathrm{AoI}^\mathrm{Rx},$ and $ \mathrm{AoII}$.

\rev{Deep reinforcement learning is not considered here, since the targeted S-IoT devices are strongly resource-constrained and such methods typically require a costly training phase with long convergence times.}

\subsection{Energy Measurements}
\label{sec:energy_measurements}

For the simulations, we assume energy consumption values corresponding to the real energy consumption used by a setup described in Sec. \ref{sec:implementation}.
To obtain these values, the energy consumption of the sensor node is measured using the Power Profiler Kit 2 (PPK2) by Nordic Semiconductor \cite{NordicPPK2}.

The device is tested during deep-sleep, and during the wake-up phase, the sensing phase, the transmission/listening phase, and the idle phase. 
\rev{In each case, we measure the average current $I_\text{avg}$ drawn from a $5$\,V supply.
Here, $I_\text{avg}$ corresponds to
\begin{equation}
\color{tudablue}
I_\text{avg} = \frac{1}{T_\text{meas}} \int_0^{T_\text{meas}} i(t)\mathrm{d}t,
\end{equation}
where $T_\text{meas}$ is the duration of the measurement interval and $i(t)$ is the current drawn at time $t$.
$I_\text{avg}$ is then multiplied by the supply voltage to obtain the mean power draw $P$ in Watt:
\begin{equation}
\color{tudablue}
P = U \cdot I_\text{avg},
\end{equation}
where $U=5V$.}

Table \ref{tab:power_measurements} summarizes the measured current and corresponding energy consumption for each mode. All measurements were performed in a stable laboratory environment.
\begin{table}[!t]
\centering
\footnotesize
\begin{tabular}{lccc}
\hline
Parameter & $I_\text{avg}$ [mA] & $P$ [W] \\
\hline
$P^\mathrm{d}$ & 0.97338   & 0.00487  \\
$P^\mathrm{w}$ & 38.25     & 0.19125  \\
$P^\mathrm{a}$ & 101.51    & 0.50755  \\
$P^\mathrm{i}$ & 27.59     & 0.13795  \\
$P^\mathrm{s}$ & 28.69     & 0.14345  \\
\hline
\end{tabular}
\caption{Measured current and energy consumption for each operation mode.}
\label{tab:power_measurements}
\end{table}

\subsection{Simulation Results}
\label{sec:simulation_results}

To evaluate the performance of \sysname, we first conduct a series of numerical experiments covering diverse system configurations.
The corresponding outcomes are presented in Fig. \ref{fig:specific} to Fig. \ref{fig:general}.
Results for \gls{got} \textbf{A} are reported in Fig. \ref{fig:specific} and Fig. \ref{fig:specific_prob}. 

\rev{Here, we assume that the device measures temperatures. 
As described in Section III, the temperature values are quantized into discrete process states.
\gls{got} \textbf{A} models penalties for a frost-watch application in which the \gls{iot} device must reliably detect when the temperature drops below the freezing point.
In the considered setup, we use a resolution of $1\,^\circ\mathrm{C}$. 
We further assume a that temperature change by plus or minus one degree with a constant probability of $0.01$. 
For \gls{got} \textbf{A}, all states below $0\,^\circ\mathrm{C}$ are classified as critical, while states at or above $0\,^\circ\mathrm{C}$ are non-critical. 
Consequently, cases where $X_t < 0\,^\circ\mathrm{C}$ while $X_t^{\mathrm{Rx}} \ge 0\,^\circ\mathrm{C}$ are penalized substantially more strongly than other inconsistencies, since the receiver misses a freezing condition. 
All other mismatches, including mismatches between two sub-zero temperatures, are still captured through the state mismatch and the AoII, but with the smaller linear scaling described above.}

The results for \gls{got} \textbf{B} are presented in Fig. \ref{fig:general_drop_rate} to Fig. \ref{fig:general_energy_weight}. 
Here, we assume that the device measures light intensities with variable state change rates derived from our own measurements in the same fashion as described above for the delays in $\mathcal{C}$.
In all figures, we show a shadow behind the mean values including $50\%$ of all outcomes ranging from the $0.25$-quantile to the $0.75$-quantile of all outcomes.

Fig. \ref{fig:specific_drop_rate} illustrates the average cost $c_\mathrm{avg}^\pi$ for different data channel erasure rates $C_\mathrm{erasure}^\mathrm{data}$ using \gls{got} \textbf{A}.
The erasure rate is held constant during each evaluation.
\sysname outperforms all baselines across the entire range.
At a low erasure rate of $0.01$, \sysname incurs a cost of $0.02$. \textit{Q-Learning} results in a cost of $0.07$, whereas \textit{AlwaysTransmit} and \textit{Threshold} result in significantly higher costs of $0.23$.

Under more challenging conditions such as $C_\mathrm{erasure}^\mathrm{data}=0.75$, the performance gap widens, with \sysname maintaining a cost of $0.06$ in contrast to $0.55$ for \textit{Q-Learning} and $2.38$ for \textit{Threshold}.

Fig. \ref{fig:specific_energy_weight} presents the average cost as a function of the energy weight $w_\mathrm{e}$, which is varied from $2^{-3}$ to $2^5$.
Across all regimes, \sysname achieves the most favorable tradeoff between \valname degradation and energy consumption.

At low energy emphasis ($w_\mathrm{e}=2^{-3}$), \sysname incurs a cost of $0.006$, whereas \textit{AlwaysTransmit} and \textit{Threshold} yield $0.03$ and $0.04$, respectively.
At $w_\mathrm{e}=2^0$, \sysname remains low at $0.02$, in contrast to \textit{Threshold}'s $0.23$.
For the extreme case $w_\mathrm{e}=2^5$, \sysname incurs a cost of $0.33$, remaining significantly below \textit{Q-Learning} ($0.43$) and \textit{Threshold} ($4.00$).
This demonstrates effective adaptation under increasing energy penalties.

Fig. \ref{fig:specific_prob} shows the impact of varying state change rates $p_\mathcal{X}$ on the average cost.
Here, we assume that the probability of the process to stay in the current state is $1-p_\mathcal{X}$ and that the probability to change to an adjacent state is given by $p_\mathcal{X}$ divided by the number of adjacent states.

As $p_\mathcal{X}$ increases from $2^{-9}$ to $2^{-2}$, \sysname retains the lowest cost across all settings.
At the lowest transition rate, \sysname yields a cost of $0.01$, compared to $0.02$ for \textit{Q-Learning} and $0.05$ for \textit{Threshold}.
As state change rates increase, energy consumption rises for adaptive schemes.
At $p_\mathcal{X}=2^{-2}$, \sysname results in a cost of $0.12$, while the costs incurred by \textit{Q-Learning} and \textit{Threshold} increase to $0.60$ and $1.47$, respectively.

In contrast, the outcome of \textit{AlwaysTransmit} remains unaffected, while \textit{NeverTransmit} and \textit{Random} occasionally benefit from high volatility due to the process returning to its initial state, which results in lower costs from \valname degradation.

Fig. \ref{fig:general_drop_rate} provides the results for $N=100$ runs, where in each run, we randomly choose one of three randomly generated \glspl{got} using the scheme \gls{got} \textbf{B} with $v=0.1, 0.2, 0.3$ under increasing $C_\mathrm{erasure}^\mathrm{data}$.
The overall trends remain consistent with those observed for \gls{got} \textbf{A}, with \sysname yielding the lowest average costs throughout.
At $C_\mathrm{erasure}^\mathrm{data}=0.01$, the cost is $0.07$, while \textit{Q-Learning} and \textit{Threshold} yield $0.31$ and $0.69$, respectively.
At $0.75$, \sysname remains at $0.28$, outperforming \textit{Q-Learning} ($1.03$) and \textit{Threshold} ($1.96$).

Fig. \ref{fig:general_energy_weight} reports results for varying $w_\mathrm{e}$ using \gls{got} \textbf{B}.
The behavior for randomly generated \glspl{got} mirrors that seen for \gls{got} \textbf{A}.
At $w_\mathrm{e}=2^{-3}$, \sysname yields a low cost of $0.02$, compared to \textit{AlwaysTransmit} ($0.03$) and \textit{Threshold} ($0.28$).
For $w_\mathrm{e}=2^0$, \sysname continues to perform well at $0.07$, while baseline costs grow steeply.
At $w_\mathrm{e}=2^5$, the cost incurred by \sysname increases to $0.67$, still undercutting \textit{Q-Learning} ($0.80$) and \textit{Threshold} ($2.94$).

\subsection{Experimental Results}
\label{sec:experimental_results}
In Fig. \ref{fig:experiments}, we present the average cost obtained by the evaluated strategies for experiments of $24$ hours for each strategy and three fixed instances of \gls{got} \textbf{B}.

Experiments for the \textit{NeverTransmit} strategy effectively correspond to only measuring the deep-sleep energy of the device, which is why we exclude them here.

The results show that \sysname achieves the lowest total cost among all compared approaches at $0.39$ compared to $0.88$, $0.96$, $3.54$, and $5.26$ for \textit{AlwaysTransmit}, \textit{Threshold}, \textit{Q-learning}, and \textit{Random}, respectively. 

These results differ from those in the simulated environment for several reasons:

First, the number of runs per strategy is lower due to the significantly higher runtime needed per experiment compared to simulations.
With fewer runs, the estimated performance becomes noisier because there is less averaging, so results can shift more due to chance and outliers.

Secondly, the difference is also due to higher and more volatile channel erasure rates, which are assumed to be fixed in the simulations.
This results in a more challenging environment for \sysname and \textit{Q-learning}.
Periods of outages lead to higher standard deviations, which becomes apparent, for example, when comparing the \valname degradation cost of \textit{AlwaysTransmit}, which is even higher than that of \sysname, as continuous synchronization is prevented by outages.

\textit{AlwaysTransmit} still exhibits low \valname degradation cost ($0.68$) compared to reference strategies.
This behavior is accompanied by the highest energy cost ($0.20$) among the evaluated approaches, which results in a higher total cost than our \sysname approach.
The energy cost for \textit{AlwaysTransmit} are more than four times higher than those for \sysname.

\textit{Random} and \textit{Threshold} achieve lower energy costs of approx. $0.01$.
However, this leads to higher \valname degradation costs ($5.25$ and $0.95$) due to the increased \gls{aoii} at the receiver, which dominates their total cost.

\textit{Q-learning} converges too slowly to yield competitive performance within the considered time horizon, even after optimizing the state space to accelerate its learning.
While it reduces \valname degradation costs compared to \textit{Random}, it is not able to reduce the cost as much as \sysname, \textit{AlwaysTransmit} or \textit{Threshold}.

\sysname finds a better tradeoff between energy consumption and \valname degradation cost than all the baseline schemes.

\section{DISCUSSION}
\label{sec:discussion}

\rev{Overall, our results suggest that \sysname can provide meaningful performance gains in the considered setting, although the current model and evaluation still have some limitations.}

\rev{The proposed approach relies on Markov models for both the observed process and the channels.
This is a deliberate modeling choice that makes the problem tractable and enables online optimization on resource-constrained devices.}

\rev{In practice, however, the observed process may exhibit temporal dependencies or non-stationary behavior that are not fully captured by a stationary Markov chain.
In principle, such processes can still be represented in a Markovian form by choosing a more expressive state definition, but this may lead to a very large state space.}

\rev{A similar limitation applies to the channel model. 
In our formulation, the heterogeneous end-to-end satellite link is represented as a single Markov channel. 
This provides a compact abstraction of time-varying delay and packet erasure behavior, but it does not resolve all physical channel effects or all link-specific dynamics individually. 
More detailed channel models could improve prediction accuracy in some scenarios, although it remains unclear how much this additional modeling effort would improve the resulting decisions about deep sleep durations in practice.}

\rev{Another important limitation is related to computational complexity. 
PSBO has to be executable on small devices.
While we already reduce implementation overhead, for example by storing the GoT metric as a function instead of a full tensor, the complexity still grows polynomially with the size of the process state space and the age cap $M$.
Therefore, the current approach is most suitable for state spaces with a moderate number of states.
Finer quantization or more expressive state representations may improve modeling accuracy, but they also increase the computational burden at the \gls{iot} device.}

\rev{Finally, the current experimental results should be seen as an initial validation rather than a complete assessment. Our simulations and hardware experiments demonstrate the potential of the approach in the considered setting, but broader validation across different observed processes, channel conditions, and hardware platforms would strengthen the conclusions. 
In addition, the present work focuses on a single sender-receiver pair, while multi-device scenarios with interacting transmissions remain an important direction for future work.}

\section{CONCLUSIONS}
\label{sec:conclusions}
In this work, we introduced \sysname, a novel algorithm for deep-sleep scheduling in \gls{siot}, which optimizes the tradeoff between energy consumption at the \gls{iot} device and \valname degradation at the receiver.
By using the \gls{got} metric, \sysname provides application-specific scheduling decisions that are aware of the usefulness of transmitted information. 

We formulated a corresponding \gls{mdp} and designed \sysname to predict future costs under uncertain channel and process dynamics.
Extensive numerical simulations demonstrated that \sysname consistently outperforms baseline methods, including a threshold-based agent and Q-learning, across a wide range of channel conditions, energy weightings, and process dynamics. 
We additionally verified the applicability of our approach and its superior performance under realistic conditions using an \gls{siot} testbed. 

Our results show the potential of the \gls{got} metric in the \gls{iot}, especially when energy consumption is a critical concern.

\bibliographystyle{IEEEtran}
\bibliography{resources/biblio}

\end{document}